\documentclass[floatfix,aps,prd,preprint,eqsecnum,tightenlines,nofootinbib,showpacs]{revtex4}
\usepackage{graphicx,epsf,amsmath}
\usepackage{subfigure}
\usepackage{relsize}
\usepackage{multirow}
\usepackage{array}
\usepackage{verbatim}

\newif\ifdraft
\drafttrue
\newif\ifpreprint
\preprinttrue

\def\fig#1{Fig.~{\ref{#1}}}
\def\Fig#1{Fig.~{\ref{#1}}}
\def\figs#1#2{Figs.~{\ref{#1}} and~{\ref{#2}}}
\def\Figs#1#2{Figs.~{\ref{#1}} and~{\ref{#2}}}
\def\sect#1{Sect.~{\ref{#1}}}

\def\tab#1{Table.~{\ref{#1}}}
\def\Tab#1{Table.~{\ref{#1}}}
\def\tabs#1#2{Tables~{\ref{#1}} and~{\ref{#2}}}

\def\e{\epsilon}
\def\eps{\epsilon}

\def\pol{\varepsilon}
\def\f{\tilde f}
\def\n{\tilde n}
\def\V{{\rm V}}

\def\tab#1{Table~{\ref{#1}}}
\def\Tab#1{Table~{\ref{#1}}}

\def\spa#1.#2{\left\langle#1\,#2\right\rangle}
\def\spb#1.#2{\left[#1\,#2\right]}
\def\tree{{\rm tree}}

\def\threeloop{{\rm 3\hbox{-}loop}}
\def\fourloop{{\rm 4\hbox{-}loop}}
\def\fiveloop{{\rm 5\hbox{-}loop}}

\def\sevenloop{{\rm 7\hbox{-}loop}}

\def\eps{\epsilon}

\def\eqn#1{Eq.~(\ref{#1})}

\def\NeqZero{{{\cal N}=0}}
\def\NeqOne{{{\cal N}=1}}

\def\NeqFour{{{\cal N}=4}}
\def\NeqFive{{{\cal N}=5}}

\def\NeqEight{{{\cal N}=8}}
\def\N{{\cal N}}
\def\fourloop{{\rm 4\hbox{-}loop}}

\def\tree{{\rm tree}}

\def\threeloop{{\rm 3\hbox{-}loop}}

\newbox\charbox
\newbox\slabox
\def\s#1{{      
        \setbox\charbox=\hbox{$#1$}
        \setbox\slabox=\hbox{$/$}
        \dimen\charbox=\ht\slabox
        \advance\dimen\charbox by -\dp\slabox
        \advance\dimen\charbox by -\ht\charbox
        \advance\dimen\charbox by \dp\charbox
        \divide\dimen\charbox by 2
        \raise-\dimen\charbox\hbox to \wd\charbox{\hss/\hss}
        \llap{$#1$} }}

\begin{document}

\ifpreprint
UCLA/14/TEP/106 \hfill $\null\hskip 4cm \null$  \hfill
\fi

\title{Enhanced Ultraviolet Cancellations in $\NeqFive$ Supergravity\\
 at Four Loops}

\author{Zvi~Bern${}^a$, Scott~Davies${}^a$ and Tristan Dennen${}^b$}
\affiliation{
${}^a$Department of Physics and Astronomy\\
University of California at Los Angeles\\
Los Angeles, CA 90095-1547, USA \\$\null$ \\
${}^b$Niels Bohr International Academy and Discovery Center\\
The Niels Bohr Institute\\
Blegdamsvej 17,\\
DK-2100 Copenhagen
 \O, Denmark
\\
$\null$
\\
$\null$
\\
}

$\null$
\begin{abstract}
\vskip .5 cm 
We show that the four-loop four-point amplitudes of
$\NeqFive$ supergravity are ultraviolet finite in four dimensions,
contrary to expectations based on supersymmetry and duality-symmetry
arguments.  We explain why the diagrams of any covariant local
formalism cannot manifestly exhibit the necessary cancellations for
finiteness but instead require a new type of ultraviolet
cancellation that we call an ``enhanced cancellation''.  We also show
that the three-loop four-point amplitudes in $\NeqFour$ and $\NeqFive$
supergravity theories display enhanced cancellations.  To construct
the loop integrand, we use the duality between color and kinematics.
We apply standard methods for extracting ultraviolet divergences in
conjunction with the {\tt FIRE5} integral reduction program to arrive
at the four-loop results.
\end{abstract}

\pacs{04.65.+e, 11.15.Bt, 11.25.Db, 12.60.Jv \hspace{1cm}}

\maketitle

\section{Introduction}

Quantum field theories of gravity are nonrenormalizable by power
counting Feynman diagrams. This leads to the widely held belief that
all unitary gravity field theories must be ultraviolet divergent at
some loop order.  Indeed, no known symmetry is powerful enough to
render such theories ultraviolet finite.  On the other hand, recent
years have made it abundantly clear that scattering amplitudes contain
hidden symmetries and new structures beyond those expected from
Lagrangians.  While these are not yet fully understood, they can have
profound consequences on ultraviolet properties.

In this paper, we identify a new class of multiloop ultraviolet cancellations
that go beyond the ones established by standard-symmetry arguments.
We call these {\it enhanced ultraviolet cancellations}.  These are
defined as cancellations that cannot be displayed term-by-term in any
local covariant diagrammatic formalism.  By such a formalism we mean
that the poles in the diagram integrands are simply the standard
Feynman propagator ones.  Using the maximal cut conditions, as defined
in Ref.~\cite{MaxCuts}, we can identify terms unique to a given
diagram which we can then power count. To illustrate enhanced
cancellations, we use previous three- and four-loop calculations in
$\NeqFour$ supergravity~\cite{ThreeLoopN4,FourLoopN4}, as well as new
calculations in $\NeqFive$ supergravity performed here.

The study of the ultraviolet properties of gravity theories has a rich
history, starting with the seminal work of 't~Hooft and
Veltman~\cite{HooftVeltman}.  They showed that pure Einstein gravity
is finite at one loop, but divergent with the addition of matter,
a point on which other papers elaborated as well~\cite{Matter}.
Goroff and Sagnotti later showed that at two loops, pure
Einstein gravity diverges~\cite{GoroffSagnotti}.  With the addition of
supersymmetry, the ultraviolet behavior tends to improve: Pure
ungauged supergravities are known to have no divergences prior to
three loops~\cite{SupergravityArguments}. However, the consensus
reached from studies in the 1980s was that all supergravity theories
would likely diverge at the third loop order (see, for example,
Ref.~\cite{HoweStelleReview}), though one can raise the loop
order with additional assumptions~\cite{GrasaruSiegel}.

The complexity of gravity theories makes it difficult to explicitly
check these expectations.  This situation was ameliorated by the
advent of the unitarity method~\cite{UnitarityMethod,BDDPR}, which
makes it possible to directly determine ultraviolet properties of
gravity theories at high loop orders.  More recently, a new constraint
on gauge-theory and gravity amplitudes has been introduced---the
duality between color and kinematics found by Carrasco, Johansson and
one of the authors (BCJ)~\cite{BCJ,BCJLoop}---allowing additional new
nontrivial computations to be carried out.

For maximally supersymmetric supergravity (in $D=4$ this is
$\NeqEight$ supergravity)~\cite{N8Supergravity}, explicit calculations
show that four-point amplitudes are finite at three loops for
dimensions $D<6$~\cite{ThreeLoopN8,CompactThree} and at four loops for
dimensions $D<11/2$~\cite{GravityFour}. In $D=4$, these ultraviolet
cancellations were subsequently understood to follow from
supersymmetry and the $E_{7(7)}$ duality symmetry of $\NeqEight$
supergravity~\cite{SevenLoopGravity, VanishingVolume}. A purely
supersymmetric explanation has also been developed by Bj\"{o}rnsson and Green~\cite{BjornssonGreen} using a field-theory version of
the Berkovits pure spinor formalism~\cite{Berkovits}.  The current
consensus based on symmetry considerations is that a $D^8 R^4$
counterterm is valid under all standard symmetries, leading to the
expectations of a seven-loop divergence in $D=4$ and a five-loop
divergence in $D=24/5$.

While technical difficulties have prevented the $\NeqEight$
supergravity expectations from being confronted by calculation, there
is now evidence that implies even better behavior in this case than that
suggested by standard-symmetry arguments: Similar argumentation in
half-maximal supergravity leads to predictions of valid counterterms
in cases where no divergences exist.  In particular, at three
loops, half-maximal $\NeqFour$ supergravity~\cite{N4Sugra} is
ultraviolet finite in four dimensions~\cite{ThreeLoopN8}, while
similar supersymmetry and duality-symmetry considerations suggest that
it should diverge~\cite{VanishingVolume}.  (See Ref.~\cite{VanhoveN4}
for string-theory arguments for finiteness.)  In addition,
half-maximal pure supergravity in $D=5$ is ultraviolet finite at two
loops, again contrary to symmetry considerations~\cite{HalfMax5D}.

An important question is whether it is possible that the observed
three-loop finiteness of $\NeqFour$ supergravity in $D=4$ can be
explained using only arguments based on supersymmetry and
duality symmetry.  An attempt to find such an explanation relied on
the assumption of the existence of an appropriate non-Lorentz
covariant off-shell 16-supercharge
superspace~\cite{BHSSuperSpace1,BHSSuperSpace2}.  Not surprisingly,
the assumption carries other consequences as well: In particular, it
predicts additional finiteness conditions when
matter multiplets are added~\cite{BHSSuperSpace2} that directly
contradict subsequent explicit calculations~\cite{N4Mat}.  Three-loop
finiteness of pure $\NeqFour$ supergravity therefore remains
unexplained by standard-symmetry arguments.  Nevertheless, it remains
a key problem to understand the extent to which such arguments can
restrict divergences.

To carry out further probes of the ultraviolet properties of
supergravity theories, together with Smirnov and Smirnov, we recently
computed the four-loop four-point divergence of $\NeqFour$
supergravity in $D=4$~\cite{FourLoopN4}, finding that the theory does
diverge at four loops. Naively, this might suggest that all
supergravity theories should diverge at some sufficiently high loop
order. However, when one looks at the details of the divergence, a
rather different picture emerges: The divergence appears to be tied to
the duality-symmetry anomaly of $\NeqFour$ supergravity found by
Marcus~\cite{MarcusAnomaly}.  The consequences of the anomaly on the
amplitudes of $\NeqFour$ supergravity have been described in some
detail in Ref.~\cite{RaduAnomaly}.  The role of the anomaly implies
that divergences of this type should not exist in ${\cal N} \ge 5$
supergravity, since these theories have no such analogous anomalies. 

In this paper, we identify a subset of terms in $\NeqEight$
supergravity that are ultraviolet divergent in four dimensions at
seven loops, reproducing the analysis of Bj\"{o}rnsson and
Green~\cite{BjornssonGreen} from a different perspective.  To identify
irreducible terms with poor power counting, we use maximal cuts.  The
expectation of a seven-loop divergence is also consistent with other
standard-symmetry arguments~\cite{CompactThree, GravityFour,
  SevenLoopGravity, VanishingVolume}.  A key question is whether there
are nontrivial enhanced cancellations between the divergent terms that
then make the amplitude as a whole ultraviolet finite.  Unfortunately,
the high required loop order makes it unfeasible at present to test
for the existence of enhanced cancellations in $\NeqEight$
supergravity.  Instead, here we demonstrate the presence of enhanced
cancellations in $\NeqFour$ and $\NeqFive$ supergravities, since they
are easier to work with, because the required loop order is lower.  As
we demonstrate in this paper, enhanced cancellations are responsible
for making the four-point $\NeqFour$ supergravity amplitudes finite at
three loops and $\NeqFive$ supergravity amplitudes finite at four
loops.  We also demonstrate three-loop cancellations in four-point
$\NeqFive$ supergravity amplitudes beyond those needed for finiteness.
Such cancellations are reminiscent of the types of nontrivial
cancellations noticed in certain unitarity cuts~\cite{Finiteness}.
The surprising aspect is that no covariant local diagrammatic
representation can make these results manifest.

What might be behind enhanced ultraviolet cancellations?  In a
previous paper with Huang~\cite{HalfMax5D}, using the duality between
color and kinematics, we explicitly tied the enhanced cancellations at
two loops in half-maximal supergravity in $D=5$ to corresponding
cancellations in pure Yang-Mills theory that prevent forbidden color
factors from appearing in divergences. A key feature is that the
ultraviolet cancellations occur between the planar and nonplanar
sectors of the theory.  This case is particularly simple to analyze in
detail because the supergravity amplitudes are simple linear
combinations of Yang-Mills amplitudes even after integration.
Unfortunately, the situation beyond two loops is much more complex
because different sets of integrals appear in the supergravity case
than in the corresponding gauge-theory case.

To carry out our computations, we use the same techniques as those
used for three and four loops~\cite{ThreeLoopN4,N4Mat,FourLoopN4} in
$\NeqFour$ supergravity.  Our computations make use of the many
advances in constructing integrands, including the unitarity
method~\cite{UnitarityMethod,BRY,BDDPR} and the duality between color
and kinematics~\cite{BCJ,BCJLoop}.  While nonplanar integrands
cannot be uniquely defined, they can still be integrated 
to obtain unique results. Once the integrands are
constructed, a mass is introduced as an infrared regulator.  We then
series expand in small external momenta (or equivalently large loop
momenta) to focus on ultraviolet
singularities~\cite{Vladimirov,ChetyrkinIBP}.  At four loops the
resulting vacuum integrals are nontrivial.  To deal with them we use
{\tt FIRE5}~\cite{FIRE5}, which implements the Laporta
algorithm~\cite{LaportaIntegrals}, to reduce the integrals to a basis
set.\footnote{We are grateful to Alexander and Volodya Smirnov
  for carrying out this step for us.}  The basis integrals are known
since they are identical to those used in the evaluation of the
four-loop QCD $\beta$ function~\cite{FourLoopIntegrals,Czakon}.

This paper is organized as follows. In \sect{MethodsSection}, we
summarize the methods used to carry out the calculations.  In
\sect{PowerCountingSection}, we review the results of
standard-symmetry power counting and show that power counting maximal
cuts gives identical results. Then in \sect{ThreeLoopSection}, we
exhibit the enhanced cancellations responsible for ultraviolet
finiteness of $\NeqFour$ supergravity at three
loops~\cite{ThreeLoopN4}.  We also present a new three-loop
calculation in $\NeqFive$ supergravity, pointing out that it too
exhibits enhanced cancellations beyond those needed for ultraviolet
finiteness.  In \sect{FourLoopSection}, we demonstrate that the
four-point amplitudes of $\NeqFive$ supergravity are all ultraviolet
finite and again display enhanced cancellations.  We present our
conclusions in \sect{Conclusions}.

\section{Methods}
\label{MethodsSection}

\subsection{Duality between color and kinematics}

The duality between color and kinematics and the associated gravity
double-copy property~\cite{BCJ,BCJLoop} make it simple to construct
supergravity amplitudes once corresponding gauge-theory amplitudes are
arranged into a form that makes the duality manifest.  (For a review
of this duality and its applications, see Ref.~\cite{HenrikJJReview}.)
At loop level the duality remains a conjecture, but we rely only on
explicitly constructed forms of $\NeqFour$ super-Yang-Mills amplitudes
where the duality is manifest~\cite{BCJLoop,ck4l}.

The duality between color and kinematics is 
usually formulated via graphs with only cubic vertices. 
Any
$L$-loop $m$-point gauge-theory amplitude with all particles in the
color-adjoint representation can be written 
in terms of such graphs as
\begin{equation}
 {\cal A}^{L-\rm loop}_m =  {i^L} {g^{m-2 +2L }}
\sum_{{\cal S}_m} \sum_{j}{\int \prod_{l=1}^L \frac{d^{D} p_l}{ (2 \pi)^{D}}
  \frac{1}{S_j}  \frac {n_j c_j}{\prod_{\alpha_j}{p^2_{\alpha_j}}}}\,.
\label{LoopGauge} 
\end{equation}
The sum labeled by $j$ runs over the set of distinct non-isomorphic
graphs, while the sum over ${\cal S}_m$ is over all $m!$ permutations
of external legs.  The symmetry factor $S_j$ removes over-counts
arising from automorphisms of the diagrams.  The product in the
denominator runs over all Feynman propagators of graph $j$, and the
integrals are over $L$ independent $D$-dimensional loop momenta.  The
color factor $c_j$ of graph $j$ is given by dressing every
three-vertex with a group-theory structure constant, $\f^{abc} = i
\sqrt{2} f^{abc}$, while $n_j$ is the kinematic numerator of graph $j$
depending on momenta, polarizations and spinors.  So far this
representation involves nothing more than absorbing contact-term
contributions into graphs with only cubic vertices by multiplying and
dividing by appropriate propagators.

\begin{figure}
\includegraphics[scale=.7]{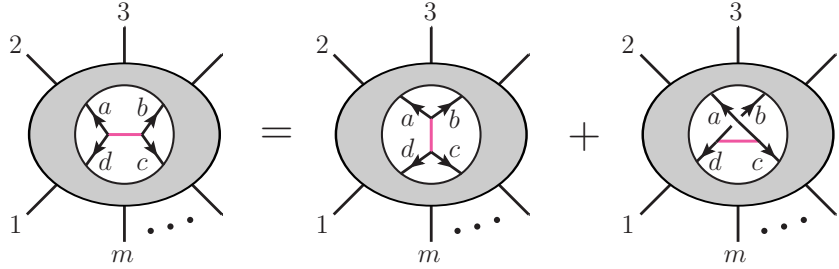}
\caption{The basic loop-level Jacobi relation for either color or
  numerator factors given in \eqn{BCJDuality}.  The basic identity
  can be embedded in a diagram at any loop order. }
\label{GeneralJacobi}
\end{figure}

The nontrivial part is the all-loop-order conjecture that there exists
a form of gauge-theory amplitudes where kinematic numerators satisfy
the same algebraic relations as color factors. These are known as
BCJ representations of amplitudes.  For the theories we discuss in
this paper, this amounts to imposing the same Jacobi identities on the
kinematic numerators as those satisfied by adjoint-representation
color factors:
\begin{equation}
c_i = c_j - c_k \;  \Rightarrow \;  n_i = n_j - n_k \,,
\label{BCJDuality}
\end{equation}
where the indices $i,j,k$ denote the diagram to which the color
factors and numerators belong.  The basic Jacobi identity is
illustrated in \fig{GeneralJacobi} embedded in an arbitrary
diagram.  The numerator factors are also required to have the same
antisymmetry properties as color factors.  In general, the duality
relations (\ref{BCJDuality}) work only after appropriate nontrivial
rearrangements of the amplitudes. 

Remarkably, we can obtain corresponding gravity loop integrands simply
by replacing color factors in a gauge-theory amplitude by kinematic
numerators of a second gauge-theory amplitude where the duality is
manifest~\cite{BCJ,BCJLoop}:
\begin{equation}
c_i \rightarrow \n_i\,.
\end{equation}
Putting in the appropriate gravitational coupling gives us 
the double-copy form of gravity amplitudes,
\begin{equation}
{\cal M}^{L-\rm loop}_m =  {i^{L+1}} {\Bigl(\frac{\kappa}{2}\Bigr)^{m-2+2L}} \,
\sum_{{\cal S}_m} \sum_{j} {\int \prod_{l=1}^L \frac{d^{D} p_l}{(2 \pi)^{D}}
 \frac{1}{S_j} \frac{n_j \n_j}{\prod_{\alpha_j}{p^2_{\alpha_j}}}} \,.
\hskip .7 cm 
\label{DoubleCopy}
\end{equation}
Only one of the two sets of numerators $n_j$ or $\n_j$ needs to
satisfy the duality relation
(\ref{BCJDuality})~\cite{BCJLoop,BCJSquare}.  We note that at tree
level ($L=0$), the duality encodes the Kawai-Lewellen-Tye~\cite{KLT}
relations between gauge-theory and gravity amplitudes, as well as
nontrivial relations between color-ordered gauge-theory partial
amplitudes~\cite{BCJ}.

\subsection{Construction of $\NeqFive$ supergravity amplitudes}
\label{N5ConstructionSubsection}

In this paper we construct the three- and four-loop four-point
$\NeqFive$ supergravity integrands using the procedure presented
above.  We do so by starting with an $\NeqOne$ super-Yang-Mills
integrand and then replacing the color factors with the BCJ forms of
kinematic numerators of $\NeqFour$ super-Yang-Mills theory given in
Refs.~\cite{BCJLoop,ck4l}.  Similar constructions of less-than-maximal
supergravity amplitudes are found in Refs.~\cite{Camille, ThreeLoopN4,
  HalfMax5D, N4Mat, FourLoopN4}.  We express
\begin{equation}
(\NeqFive \hbox{ sugra} )  \;\; : \;\;
 (\NeqFour \hbox{ sYM}) \otimes (\NeqOne \hbox{ sYM})\,,
\end{equation}
 where ``sugra'' and ``sYM'' are shorthands for, respectively,
supergravity and super-Yang-Mills theory.  We further decompose the
$\NeqFive$ amplitudes into a direct sum,
\begin{align}
& (\NeqFive \hbox{ sugra} )  \;\; : \; \; (\NeqFour \hbox{ sYM}) \otimes 
 (\NeqZero \hbox{ sYM}) \nonumber\\ 
& \hskip 3.3 cm 
\oplus 
(\NeqFour \hbox{ sYM}) \otimes
 \Bigl((\NeqOne \hbox{ sYM}) \ominus (\NeqZero \hbox{ sYM})\Bigr) \,,
\label{NeqOneSusyDecomp}
\end{align}
where ``$\NeqZero \hbox{ sYM}$'' refers to ordinary pure
nonsupersymmetric Yang-Mills theory.  The first term in the direct sum is
the pure $\NeqFour$ supergravity amplitude, while the second
term is the difference between the $\NeqFive$ and $\NeqFour$
supergravity amplitudes.  The second term comes from taking the
diagrams of pure $\NeqOne$ super-Yang-Mills, subtracting out the
pure-gluon part, and then replacing the color factors with the BCJ
numerators of $\NeqFour$ super-Yang-Mills theory. 
On the $\NeqOne$ super-Yang-Mills
side, this amounts to separating the pure gluon contributions
from the contributions including also gluinos.

Following Refs.~\cite{ThreeLoopN4,N4Mat,FourLoopN4}, we use ordinary
Feynman diagrams for the $\NeqZero$ and $\NeqOne$ super-Yang-Mills
integrands.  While this might seem to be a poor starting point given
the complexity of such diagrams, the BCJ construction
(\ref{DoubleCopy}) ensures that only a small fraction of diagrams
actually contribute.  Whenever an $\NeqFour$ super-Yang-Mills diagram
vanishes, we do not need to evaluate corresponding diagrams in
$\NeqZero$ or $\NeqOne$ super-Yang-Mills theory.  At one, two, three
and four loops, the BCJ-satisfying representations of the four-point
amplitudes of $\NeqFour$ super-Yang-Mills theory have only,
respectively, 1, 2, 12 and 85 nonvanishing diagrams (up to
permutations of external legs).  This is already a remarkable
simplification, allowing the calculation to proceed.

The decomposition in Eq.~\eqref{NeqOneSusyDecomp} results in an
integrand where the $\NeqFour$ supersymmetry cancellations are
manifest, but the $\NeqOne$ ones are not.  While we do not do so here,
one could simplify the $\NeqFive$ supergravity integrand to make
cancellations from all supersymmetries manifest.  This could be
accomplished by using the unitarity method to systematically move
terms between diagrams, subject to maintaining the unitarity cuts.
However, as we shall see below, no covariant local representation
exists either in $\NeqFour$ or $\NeqFive$ supergravity that makes
manifest the complete set of ultraviolet cancellations that we find.
Because of this, there is no obvious way to avoid direct
integration to see the cancellations.  We consequently call such 
cancellations enhanced.

\subsection{Extraction of ultraviolet divergences}

Once we have an integrand, the next step is to extract the ultraviolet
divergences.  The procedure that we use has been described in some
detail in Ref.~\cite{N4Mat}, so here we only briefly summarize it.
To deal with potential ultraviolet divergences we use dimensional
reduction~\cite{DimRed}.  Rather than evaluate integrals with their
full momentum dependence, it is much simpler to series expand the
integrands prior to integration in order to pick up only the desired
ultraviolet divergences~\cite{Vladimirov}.  This procedure introduces
new unphysical infrared singularities beyond the standard ones, so one
needs an infrared cutoff to separate the ultraviolet divergences from
the infrared ones.

An especially good choice for regulating infrared singularities is to
introduce a uniform mass into all Feynman propagators prior to
expanding in external
momenta~\cite{ChetyrkinUniformMass,ChetyrkinThreeLoop}. For the cases
we study in this paper, where there are no lower-loop divergences, the
subdivergences should all cancel amongst themselves with the use of
this regulator.  The uniform mass regulator therefore greatly
simplifies the computation since we do not need to compute
subdivergences. We have, however, performed extensive checks
confirming that they cancel as expected.  We note that if the mass
regulator were introduced later in the calculation, for example after
the expansion in external momenta and tensor integral simplifications,
it would ruin the cancellations of subdivergences between different
integrals.  One would then need to include all subdivergence
subtractions to properly remove the regulator dependence, greatly
complicating the calculation.

The procedure results in a large number of vacuum integrals.  At three
loops, evaluating the integrals is
straightforward~\cite{ChetyrkinThreeLoop, ThreeLoopN4}, but at four
loops it is a more serious challenge.  To deal with this, we use the
{\tt FIRE5} program~\cite{FIRE5}, which is a highly-efficient
implementation of integration-by-parts relations \cite{ChetyrkinIBP}
using the Laporta algorithm~\cite{Laporta}.  It allows us to write
down any given integral as a linear combination of a small number of
so-called master integrals.  In our four-loop calculation, the
reduction to master integrals is especially nontrivial due to the high
powers of numerator loop momenta that occur in gravity.  Earlier
related calculations already determined the four-loop vacuum master
integrals~\cite{FourLoopIntegrals, LaportaIntegrals, Czakon}; we use
the master-integral basis and values given in Ref.~\cite{Czakon}.


\section{Power counting}
\label{PowerCountingSection}

\subsection{Review of standard-symmetry power counting}

\begin{table}[t]
\begin{center}
\begin{tabular}[t]{||l|c|c||}
\hline
\; Theory & \; Counterterm \; & Loop Order\\
\hline\hline
$\; D=4, \;Q = 32,\; \NeqEight\;$ & ${\cal D}^8 R^4$ & 7 \\
\hline
$\; D=4,\; Q = 16,\; \NeqFour\;$ & $R^4$ & 3 \\
\hline
$\; D=4,\; Q = 20, \;\NeqFive\;$ & ${\cal D}^2R^4$ & 4 \\
\hline
$\; D=24/5,\; Q = 32\;$ & ${\cal D}^8R^4$ & 5 \\
\hline
$\; D=5,\; Q = 16\;$ & $R^4$ & 2 \\
\hline 
\end{tabular}
\end{center}
\caption{Selected valid counterterms based on supersymmetry and
  duality-symmetry considerations~\cite{SevenLoopGravity,BjornssonGreen,
    VanishingVolume,HalfMax5D, BHSSuperSpace2, N4Mat}. 
   $Q$ is the number of supercharges.  }
\label{CounterTermsTable}
\end{table}

The restrictions supersymmetry and duality symmetry impose on
counterterms have been studied in great detail over the years.  The
most recent power-counting predictions based on symmetry
considerations are collected in \tab{CounterTermsTable}.  In $D=4$,
apparently valid counterterms exist at loop orders $L=7$ in
$\NeqEight$ supergravity~\cite{SevenLoopGravity,BjornssonGreen,
  VanishingVolume}, $L=3$ in $\NeqFour$
supergravity~\cite{VanishingVolume}, and $L=4$ in $\NeqFive$
supergravity~\cite{VanishingVolume}.  By increasing the space-time
dimensions, one can also lower the loop order at which a potential
counterterm can correspond to a divergence.  For example, in $D=24/5$,
maximal 32-supercharge supergravity has a valid five-loop
counterterm~\cite{BjornssonGreen}.  Similarly, half-maximal
16-supercharge supergravity in $D=5$ has an apparently valid two-loop
counterterm~\cite{HalfMax5D, BHSSuperSpace2, N4Mat}.  As explained in
Ref.~\cite{VanishingVolume}, the counterterms listed in
\tab{CounterTermsTable} cannot be written as full-superspace
integrals, but they do appear to be valid under all known
standard-symmetry considerations.  See also
Refs.~\cite{BHSSuperSpace1, BHSSuperSpace2, N4Mat} for an attempt
to put tighter restrictions on the counterterms and the associated
difficulties with doing so.

Bj\"{o}rnsson and Green~\cite{BjornssonGreen} constructed a
first-quantized pure spinor formalism useful for power counting.  Their
formalism exposes all supersymmetry cancellations and gives an
identical power count as other recent methods, including those that
account for duality symmetry~\cite{SevenLoopGravity, VanishingVolume}.
Their results imply that unless there are some extra nonstandard
cancellations beyond those implied by supersymmetry, $\NeqEight$
supergravity will diverge at five loops in $D=24/5$ and at seven loops
in $D=4$, corresponding to the first and fourth rows of
\tab{CounterTermsTable}.  We know that through four loops in
$\NeqEight$ supergravity, such symmetry-based predictions match the
explicitly computed critical dimensions where a divergence first
appears~\cite{GSB,BDDPR, CompactThree, BCJLoop, ck4l}.  A key question
is whether this pattern continues or whether there are cancellations
beyond the well-understood ones.

While it is not currently feasible to answer this for $\NeqEight$
supergravity, we can answer it for $\NeqFour$ and $\NeqFive$
supergravity.  From previous work~\cite{ThreeLoopN4}, we already know that the
three-loop $R^4$ counterterm of $\NeqFour$ supergravity in $D=4$ listed
on the second row of \tab{CounterTermsTable} does {\it not} result in
a three-loop divergence.  Similarly, half-maximal
16-supercharge supergravity at two loops in $D=5$ is free of
divergences~\cite{HalfMax5D}. As we discuss below in some detail for
the three-loop $\NeqFour$ case, these cancellations are a nontrivial
manifestation of enhanced cancellations.  However, one may worry that
these two cases are special and not representative of a general
pattern.  In particular, $\NeqFour$ supergravity in $D=4$ has a $U(1)$
anomaly~\cite{MarcusAnomaly} that would not occur in theories with
higher supersymmetry.  The two-loop $D=5$ case also has some special
features: At two loops the BCJ kinematic numerators of maximally
supersymmetric Yang-Mills four-point amplitudes are independent of
loop momenta, implying that half-maximal supergravity amplitudes are
simple linear combinations of the corresponding pure Yang-Mills
ones. To go beyond these special cases, here we study the case of
$\NeqFive$ supergravity in $D=4$ to show that there is no divergence
associated with the counterterm listed on the third line of
\tab{CounterTermsTable}.  This case is not entangled with any known
anomaly.  Furthermore, unlike the two-loop case, the kinematic
numerators do depend on loop momenta, so the gravity integrals are
different from the corresponding Yang-Mills ones.

\subsection{Power counting maximal cuts}

In order to describe the phenomenon of enhanced cancellations, we
turn to power counting using maximal cuts.  The terms selected by a
maximal cut are a gauge-invariant set that are unique to a diagram.
Using maximal cuts, we can incorporate all supersymmetric
cancellations into supergravity power counts using the known power
counts of super-Yang-Mills theories. Because all supersymmetric
cancellations are accounted for, this gives us a power-counting method
equivalent to the one of Bj{\"o}rnsson and
Green~\cite{BjornssonGreen}.

The maximal cut of a given diagram is obtained by replacing all
propagators with on-shell conditions.  While the cut conditions set
various terms to zero, they do allow us to identify terms with poor
behavior, in some cases worse behavior than that of the full
amplitude.  Once we have selected terms using the maximal cuts, we
promote them back to Feynman integrals, making sure that the obtained
representation has the {\it minimum power count} consistent with the
cut.  If any term is then found whose ultraviolet behavior is worse
than that of the amplitude as a whole, then by definition, we have
enhanced cancellations.  Of course, some care is required to be sure
that we are using a form that has minimum power count but is also
consistent with the cut.  To be clear, we are defining enhanced
cancellations entirely by their integrand power-counting properties and not by
cancellations that appear only after integration.  
We do the power counting in $D$ dimensions, viewing
$D$ as arbitrarily large, to not include hidden relations in the cut
solutions that might lead to extra cancellations. In this way 
we maintain $D$-dimensional covariance.

We will show that the maximal cuts give power counts equivalent to the
potential counterterms in \tab{CounterTermsTable}. This should not be
too surprising given that the maximal cuts are a gauge-invariant
subset built from objects that respect all standard symmetries of the
amplitudes.  As with other power counts, the maximal cuts do not make
enhanced cancellations visible because they do not account for
nontrivial cancellations between diagrams. Indeed, at a sufficiently
high loop order, the amplitudes of {\it every} supergravity theory
necessarily have divergences in individual terms selected by the
maximal cuts.  To see a better behavior in unitarity cuts, one needs
to instead look at cuts that collect together many diagrams so as to
allow cancellations between them.  The cuts analyzed in
Ref.~\cite{Finiteness} suggesting improved all-loop behavior of the
amplitudes are examples of this.

\begin{figure}[t]
\centering
\renewcommand{\subfigcapskip}{-.05cm}
\subfigure[]{\includegraphics[clip,scale=0.45]{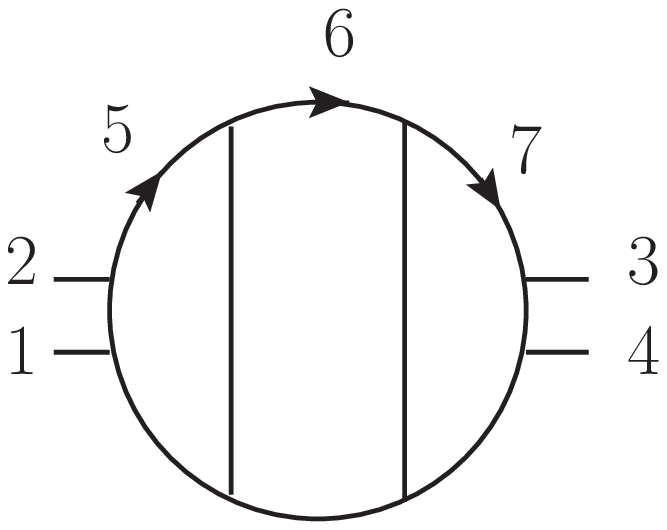}}
  \hskip .5 cm
\subfigure[]{\includegraphics[clip,scale=0.45]{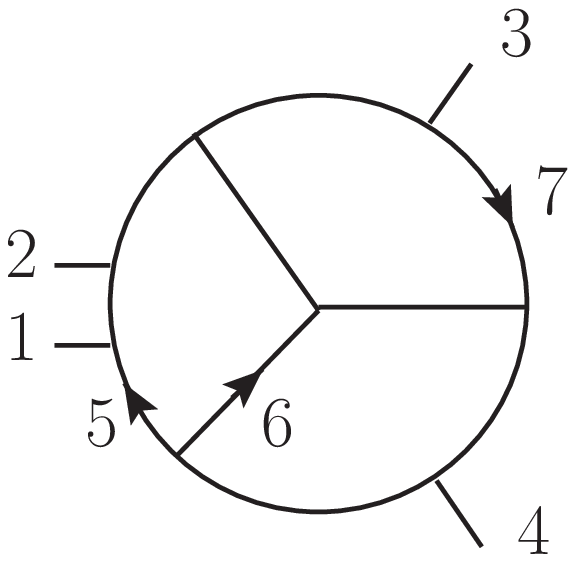}}
\caption[]{Three-loop sample diagrams for maximal-cut power counting.} 
\label{MaxCutThreeLoopFigure}
\end{figure}

As a warm-up, we first consider maximal cuts in $\NeqEight$
supergravity. We consider the diagram in
\fig{MaxCutThreeLoopFigure}(a) as a simple first example.  In
$\NeqEight$ supergravity, a kinematic numerator consistent with the
maximal cuts is given by~\cite{ThreeLoopN8}
\begin{equation}
N^{\threeloop}_{\NeqEight\ \rm sugra} = s^5 t u M_4^\tree \,,
\end{equation}
where $M_4^\tree$ is the four-point gravity tree amplitude,
and $s, t$ and $u$ are the standard four-point Mandelstam invariants.
The maximal-cut conditions have no effect on this numerator since
it is independent of loop momentum.
Counting the three $D$-dimensional loop integrals, no powers of
loop momentum in the numerator, and ten propagators gives us 
the power count,
\begin{equation}
{\cal D}^{\threeloop}_{\NeqEight\ \rm sugra} \sim \Lambda^{3D-20}\,,
\end{equation}
where $\Lambda$ is an ultraviolet cutoff.  The critical dimension
where an ultraviolet divergence first occurs is thus $D_c = 20/3$ for
the maximal-cut terms of this diagram.  However, the $\NeqEight$
three-loop amplitude also contains worse-behaved terms.  We consider
instead the diagram in \fig{MaxCutThreeLoopFigure}(b).  In $\NeqEight$
supergravity, a kinematic numerator consistent with the unitarity cuts
of this diagram is given in Ref.~\cite{CompactThree}:
\begin{equation}
N^{\threeloop}_{\NeqEight\ \rm sugra} = s^3 t u M_4^\tree (l_5 - k_4)^4\,,
\end{equation}
where the momenta correspond to the labels in the diagram.
Applying the maximal-cut conditions, we
set $l_5^2 = 0$ and obtain the minimal power-counting form, 
\begin{equation}
N^{\threeloop}_{\NeqEight\ \rm sugra}\bigr|_{\rm max.\; cut} = 
s^3 t u M_4^\tree (2 l_5\cdot k_4)^2\,.
\end{equation}
After promoting this back to the numerator of a full three-loop Feynman
integral, we count the powers of loop momenta.
Counting the three $D$-dimensional loop integrals, two powers of
loop momentum in the numerator, and ten propagators gives us an
overall power count for the diagram of
\begin{equation}
{\cal D}^{\threeloop}_{\NeqEight\ \rm sugra} \sim \Lambda^{3D +2 -20}\,.
\end{equation}
The critical dimension of this contribution is thus $D_c = 6$, which
matches the critical dimension obtained from explicit divergence
calculations~\cite{ThreeLoopN4,CompactThree}.  In this case then,
there are no enhanced cancellations. 

\begin{figure}[t]
\centering
\renewcommand{\subfigcapskip}{-.05cm}
\subfigure[]{\includegraphics[clip,scale=0.45]{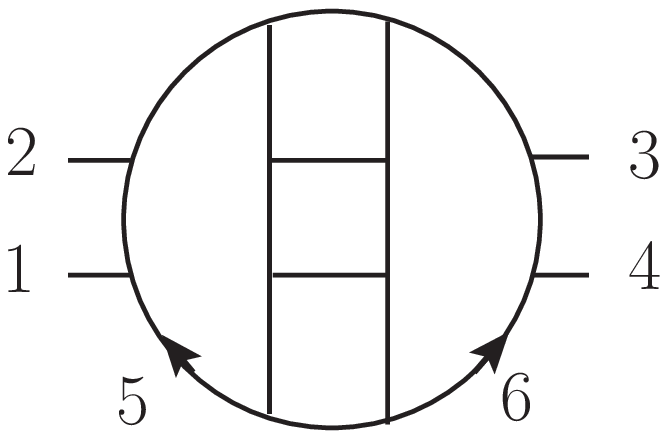}}
  \hskip .5 cm
\subfigure[]{\includegraphics[clip,scale=0.45]{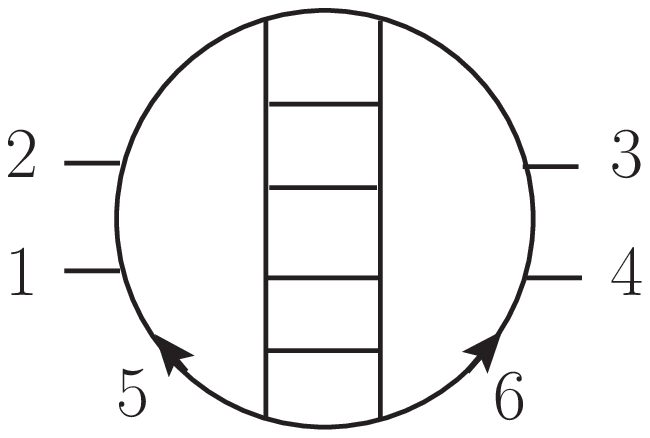}}
\vskip -.2 cm 
\caption[]{Sample diagrams for power counting maximal cuts
at five and seven loops.}
\label{MaxCutFiveSevenLoopFigure}
\end{figure}

Next we consider the maximal cut of the five-loop diagram in
\fig{MaxCutFiveSevenLoopFigure}(a).\footnote{The importance of these
  types of cuts for power counting was first pointed out by Henrik
  Johansson.}  The simplest numerator consistent with the diagram's
maximal cut in $\NeqEight$ supergravity is
\begin{equation}
N^{\fiveloop}_{\NeqEight\ \rm sugra}\bigr|_{\rm max.\; cut} = 
 s^5 t u M_4^\tree (2 l_5 \cdot l_6)^4\,,
\label{FiveLoopNumerator}
\end{equation}
where the momenta follow the labels of 
\fig{MaxCutFiveSevenLoopFigure}(a). The numerator follows from the
rung rule~\cite{BRY,BDDPR}---a rule devised to give the correct
iterated two-particle cuts---after dropping terms that vanish
with the on-shell conditions $l_i^2 = 0$.  
Promoting the maximal-cut terms back to numerators of a
Feynman integral, we have five $D$-dimensional loop integrals and sixteen
propagators. Together with the numerator (\ref{FiveLoopNumerator}),
we then have a power count,
\begin{equation}
{\cal D}^{\fiveloop}_{\NeqEight\ \rm sugra} \sim \Lambda^{5D +8 -32}\,.
\end{equation}
This gives a critical dimension $D_c = 24/5$, matching the
Bj\"{o}rnsson and Green analysis~\cite{BjornssonGreen}. After stepping
through the diagrams, this turns out to be the worst-behaved
contribution.  If it were to turn out that the critical dimension of
the full amplitude is greater than $24/5$, then by definition there
would be enhanced ultraviolet cancellations.

Similarly, we go through the same exercise for the seven-loop
diagram shown in \fig{MaxCutFiveSevenLoopFigure}(b).
 In this case, the simplest form of the numerator
consistent with the maximal cuts is
\begin{equation}
N^{\sevenloop}_{\NeqEight\ \rm sugra}\bigr|_{\rm max.\; cut}
 =  s^5 tu M_4^\tree (2 l_5 \cdot l_6)^8 \,.
\label{SevenLoopNumerator}
\end{equation}
Together with seven $D$-dimensional loop integrations and 22 propagators, we
obtain the power count,
\begin{equation}
{\cal D}^{\sevenloop}_{\NeqEight\ \rm sugra} \sim \Lambda^{7D +16 -44} \,.
\end{equation}
Thus, the critical dimension is $D_c = 4$, again in agreement with
other power-counting methods~\cite{SevenLoopGravity, BjornssonGreen,
  VanishingVolume}. 

While it is not yet technically feasible to directly study the
enhanced cancellations in $\NeqEight$ supergravity five- and
seven-loop amplitudes, we are able to study them in $\NeqFour$ and $\NeqFive$
supergravities. We therefore turn to power counting in these theories.

Consider $\NeqFour$ supergravity at three loops.  As explained in
Ref.~\cite{ThreeLoopN4}, the BCJ construction of the integrand is in
terms of the 12 diagrams displayed in \fig{ThreeLoopFigure}.  To be
concrete, we examine diagram (a) in \fig{MaxCutThreeLoopFigure} for
$\NeqFour$ supergravity.  The maximal-cut conditions on the kinematic
invariants are \def\hs{\hskip 1 cm}
\begin{align}
& l_5^2 = l_6^2 = l_7^2 = 0,  \hs l_6\cdot l_7= 0,\hs
l_5\cdot l_6= 0, \hs  k_2\cdot l_7 =  -\frac{s}{2} - k_1\cdot l_7, 
\nonumber \\
& k_3\cdot l_7 = 0, 
\hs k_2\cdot l_6 = -\frac{s}{2} - k_1\cdot l_6,\hs 
k_1\cdot l_5 = -\frac{s}{2},
\hs
k_2\cdot l_5 = 0\,.
\label{MaxCuts}
\end{align}
Applying these and taking the explicit expression for the numerator of
\fig{MaxCutThreeLoopFigure}(a) in $\NeqFour$ supergravity obtained by
the double-copy procedure, we obtain
\begin{equation}
N^{{\rm (a)}\threeloop}_{\NeqFour\ \rm sugra}\bigr|_{\rm max.\; cut}
 = -64 s^3 t \, A_{\NeqFour}^\tree \,
(\pol_1 \cdot l_5) \,( \pol_2 \cdot l_5) \,
(\pol_3 \cdot l_7) \, (\pol_4 \cdot l_7) \, (l_5 \cdot l_7)^2 
+ \cdots
\,,
\label{DiagramaNumerator}
\end{equation}
where we kept only those term with the largest powers of loop momenta.
The momentum labels are the ones shown in the figure and $
A_{\NeqFour}^\tree$ is an $\NeqFour$ super-Yang-Mills tree amplitude
depending only on the external states and momenta.  The $\pol_i$ are
polarization vectors of gluons.  As discussed in
\sect{MethodsSection}, the pure $\NeqFour$ supergravity states are
just the direct product of states of the two gauge theories.  The
displayed term in \eqn{DiagramaNumerator} is irreducible in that its
power count cannot be lowered by imposing the maximal-cut conditions
(\ref{MaxCuts}).  Since the term (\ref{DiagramaNumerator}) is uniquely
assigned to the diagram, it is a lower bound on the power count of
the diagram.  After including the three $D$-dimensional loop
integrals, eight powers of numerator loop momentum and ten
propagators, we obtain a power-counting for this diagram,
\begin{equation}
{\cal D}_{\NeqFour\ \rm sugra}^{\rm (a)\, \threeloop} 
                      \sim \Lambda^{3 D +8 - 20} \,.
\label{ThreeLoopPowerCountN4}
\end{equation}
Thus, in $D=4$ this diagram has divergent terms.  As a direct
confirmation of this power count, we integrated the irreducible
numerator in \eqn{DiagramaNumerator} after putting back the
propagators.  Indeed, it is ultraviolet divergent as indicated from
the power count.  This power count agrees with the one based on
standard-symmetry arguments~\cite{VanishingVolume}.

On the other hand, explicit calculations show that the three-loop
four-point $\NeqFour$ supergravity amplitude is
finite~\cite{ThreeLoopN4}.  Given that there are divergent terms in
\fig{MaxCutThreeLoopFigure}(a) that can only cancel against terms that
were set to zero by the maximal cut conditions or terms from other
diagrams, the finiteness of the amplitude as a whole is a prime
example of an enhanced cancellation.

The maximal-cut constraints can sometimes lower the power count of
diagrams below their true critical dimension. For example, for the
diagram in \fig{MaxCutThreeLoopFigure}(b), under the maximal-cut
conditions, all $l_i \cdot l_j$ can be made to be no worse than linear
in loop momenta.  By choosing the minimal resulting power count, this
results in an integrand that is ultraviolet finite, even after
including an extra power of loop momentum from the $\NeqFour$
super-Yang-Mills side of the double copy.  Another point is that after
integration, it may be possible to combine terms even from a single
diagram to get a finite result.  In particular, one can imagine taking
the numerator of \eqn{DiagramaNumerator} and combining it with a
judiciously chosen set of terms that vanish on the cuts to cancel the
ultraviolet divergences.  However, this is not relevant for enhanced
cancellations which are defined in terms of power counting individual
terms at the integrand level.  If even a single term in the integrand
of a single diagram has a worse power count compared to the actual
behavior of the full amplitude and the power count cannot be lowered
by maximal-cut conditions, then we have identified enhanced
cancellations.  We also note that, by using spinor-helicity, we can
set the integrated divergence resulting from the diagram in
\fig{MaxCutThreeLoopFigure}(a) to zero (i.e.,  the integration results
in terms containing $\pol_i \cdot \pol_j$ that can be set to zero by
special reference momentum choices).  Indeed, we shall do so later to
simplify various tables.  Of course, this does not change the fact
that there is a term in the integrand that has a worse power count
than the amplitude as a whole that cannot be set to zero.

The counting for $\NeqFive$ supersymmetry is similar except that one
should subtract two powers of loop momenta from each numerator
because of additional $\NeqOne$ supersymmetry cancellations 
described in Ref.~\cite{GrasaruSiegel}.  Taking this into account, 
for the diagram \fig{MaxCutThreeLoopFigure}(a), we
obtain a maximal-cut power count in $\NeqFive$ supergravity of
\begin{equation}
{\cal D}_{\NeqFive\ \rm sugra}^{\rm (a)\, \threeloop} 
                       \sim \Lambda^{3 D + 6 - 20} \,. 
\end{equation}
Thus, the critical dimension from the maximal cut of this diagram is
$D_c = 14/3 >4$, so we expect there to be no obstruction to finding a
covariant representation of $\NeqFive$ supergravity that is manifestly
ultraviolet finite in $D=4$ at this loop order.  Nevertheless, in
\sect{ThreeLoopSection} we will show that on top of the supersymmetric
cancellations, there are additional enhanced cancellations beyond
those needed for ultraviolet finiteness.

\begin{figure}[t]
\includegraphics[clip,scale=0.43]{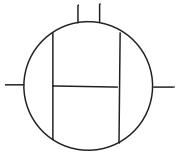}
\caption[a]{A four-loop diagram whose maximal-cut power count
  suggests that $\NeqFive$ supergravity should diverge in four
  dimensions, contrary to the behavior of the four-point amplitude as
  a whole.}
\label{MaxCutFourLoopFigure}
\end{figure}

If we repeat the same exercise at four loops for $\NeqFour$
supergravity using similar power counting on, for example, the diagram in
\fig{MaxCutFourLoopFigure}, we have the behavior, 
\begin{equation}
{\cal D}_{\NeqFour\ \rm sugra}^{\fourloop} \sim \Lambda^{4 D + 12 - 26} \,,
\end{equation}
Here we count four $D$-dimensional loop integrals, 13 propagators, 10
powers of numerator loop momenta from the Yang-Mills vertices and 2
additional powers of loop momenta from the $\NeqFour$ super-Yang-Mills
numerator.  This means that terms in the diagram in
\fig{MaxCutFourLoopFigure} have a critical dimension of $D_c = 14/4 <
4$ and that in $D = 4$ it is quadratically divergent by power counting.

If we increase the supersymmetry to $\NeqFive$ supergravity, as noted
above, the extra $\NeqOne$ supersymmetry decreases the maximal-cut
power count by two powers of loop momentum so that
\begin{equation}
{\cal D}_{\NeqFive\ \rm sugra}^{\fourloop} \sim \Lambda^{4 D + 10 - 26} \,,
\end{equation}
which corresponds to a critical dimension of $D_c = 4$.  Therefore,
based on the maximal-cut power counting, we would expect $\NeqFive$
supergravity to be logarithmically divergent at four loops.  This is
consistent with the standard-symmetry power count of
Ref.~\cite{VanishingVolume}, leading to an expected counterterm on the
third line of \tab{CounterTermsTable}.  In \sect{FourLoopSection} we
show that because of enhanced cancellations, the $\NeqFive$ four-loop
four-point amplitude is, in fact, ultraviolet finite, contrary to
these power counts.

\section{Three loops}
\label{ThreeLoopSection}

\begin{figure}[tb]
\centering
\renewcommand{\subfigcapskip}{-.1cm}
\renewcommand{\subfigbottomskip}{0.25 cm}
\hspace{0.2cm}
\subfigure[]{\includegraphics[clip,scale=0.35]{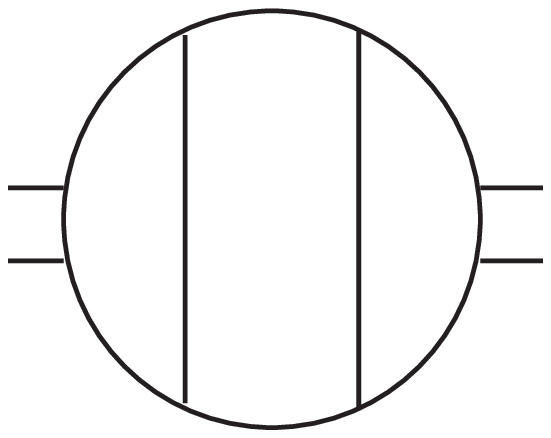}}
\hspace{0.2cm}
\subfigure[]{\includegraphics[clip,scale=0.35]{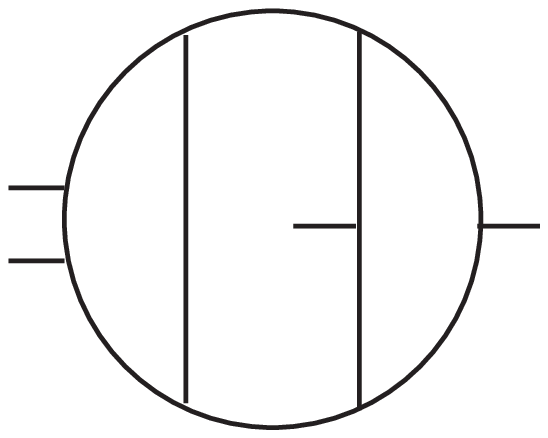}}
\hspace{0.4cm}
\subfigure[]{\hskip -.2cm\includegraphics[clip,scale=0.35]{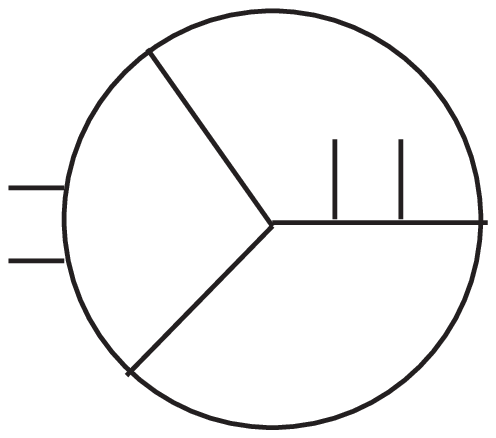}}
\hspace{0.4cm}
\subfigure[]{\includegraphics[clip,scale=0.35]{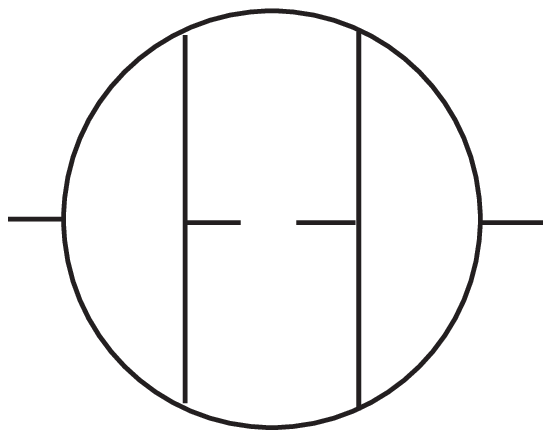}}
\\
\hspace{0.2cm}
\subfigure[]{\hskip -.2cm \includegraphics[clip,scale=0.35]{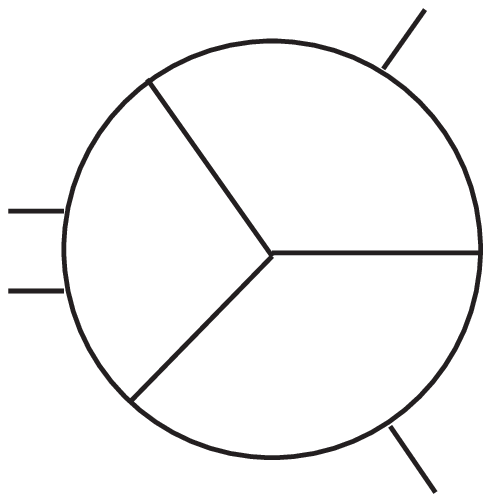}}
\hspace{0.6cm}
\subfigure[]{\hskip -.2cm \includegraphics[clip,scale=0.35]{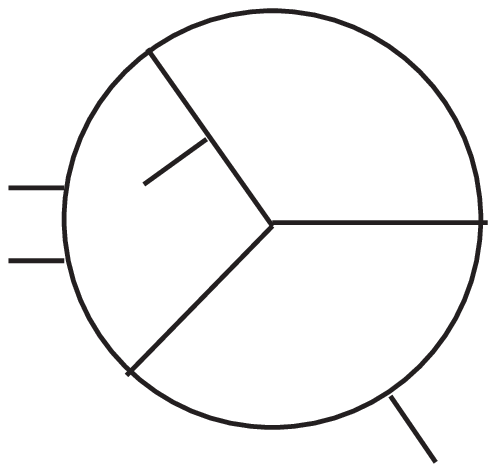}}
\hspace{0.6cm}
\subfigure[]{\hskip -.2cm \includegraphics[clip,scale=0.35]{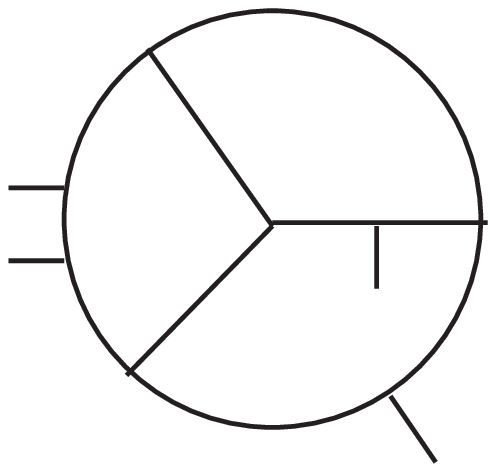}}
\hspace{0.6cm}
\subfigure[]{\includegraphics[clip,scale=0.35]{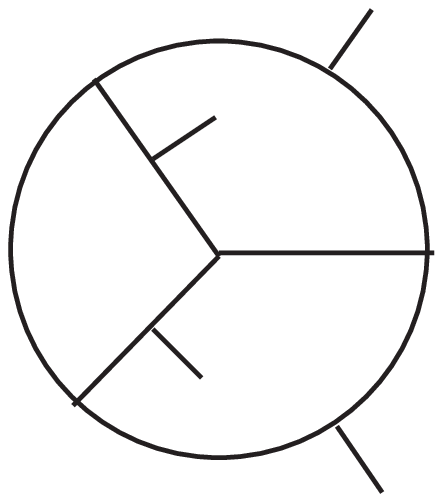}}
\\
\hspace{0.2cm}
\subfigure[]{\hskip -.2cm\includegraphics[clip,scale=0.35]{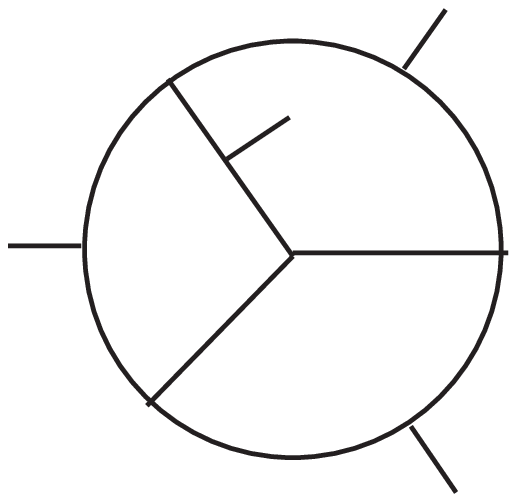}}
\hspace{0.6cm}
\subfigure[]{\hskip -.4cm\includegraphics[clip,scale=0.35]{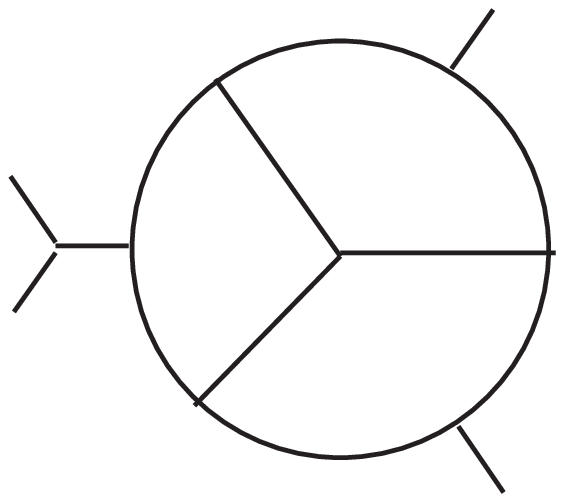}}
\hspace{0.5cm}
\subfigure[]{\hskip -.3cm\includegraphics[clip,scale=0.35]{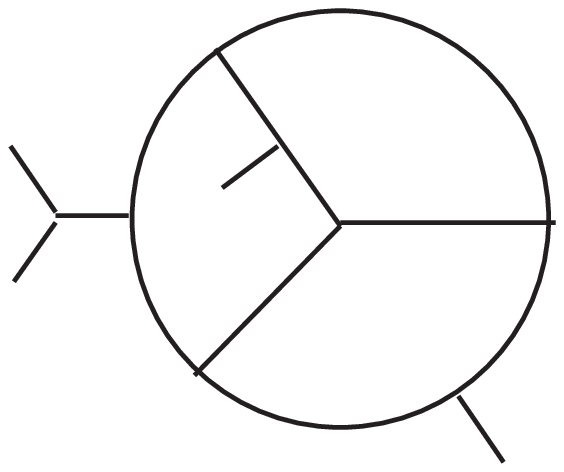}}
\hspace{0.45cm}
\subfigure[]{\hskip -.4cm\includegraphics[clip,scale=0.35]{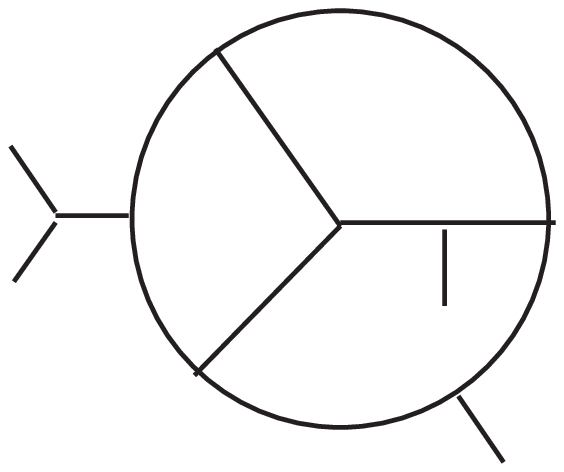}}
\caption{Contributing three-loop diagrams in $\NeqFour$ and $\NeqFive$ 
supergravity. }
\label{ThreeLoopFigure}
\end{figure}

\begin{table}[t]
  \begin{center}
  \scalebox{.9}{
\begin{tabular}[t]{||c|c||}
\hline
graph & $\vphantom{\bigg|}(\text{divergence})(4\pi)^6/(\langle 12 \rangle^2 [34]^2 stA^\tree (\frac{\kappa}{2})^8)$ \\
\hline \hline
(a)--(d) & \vphantom{\bigg|}
0 
\\ \hline
(e) & \begin{tabular}{@{}c@{}}
$\vspace{-.2cm}\vphantom{\bigg|}
\left(-\frac{77}{768}+\frac{85D_s}{768}\right)\frac{1}{\e^3}+\left(\frac{35071}{55296}-\frac{2371D_s}{6912}-\frac{D_s^2}{64}\right)\frac{1}{\e^2}+\Big[\left(-\frac{11815}{768}+\frac{4367D_s}{768}+\frac{9D_s^2}{32}\right){\rm S2} \;$ \\
$\vphantom{\bigg|}+\left(-\frac{77}{512}+\frac{85D_s}{512}\right)\zeta_2+\left(\frac{10627}{2304}-\frac{1705D_s}{576}\right)\zeta_3-\frac{170275}{110592}+\frac{1381D_s}{1536}-\frac{5D_s^2}{128}\Big]{} \frac{1}{\e} {}$
 \end{tabular} \\ \hline
(f) & \begin{tabular}{@{}c@{}}
 $\vspace{-.2cm}\vphantom{\bigg|}\left(\frac{397}{2304}-\frac{143D_s}{2304}\right)\frac{1}{\e^3}+\left(\frac{1717}{4608}+\frac{211D_s}{4608}\right)\frac{1}{\e^2}+\Big[\left(-\frac{1777}{256}+\frac{263D_s}{256}\right){\rm S2}$ \\
$\vphantom{\bigg|}+\left(\frac{397}{1536}-\frac{143D_s}{1536}\right)\zeta_2+\left(-\frac{649}{288}+\frac{69D_s}{64}\right)\zeta_3+\frac{685733}{165888}-\frac{116663D_s}{82944}\Big]\frac{1}{\e}$ 
\end{tabular} \\ \hline
(g) & \begin{tabular}{@{}c@{}} 
$\vspace{-.2cm}\vphantom{\bigg|}\left(-\frac{23}{288}-\frac{65D_s}{1152}\right)\frac{1}{\e^3}+\left(-\frac{7919}{6912}+\frac{4631D_s}{13824}+\frac{D_s^2}{192}\right)\frac{1}{\e^2}+\Big[\left(\frac{2447}{96}-\frac{2911D_s}{384}-\frac{3D_s^2}{32}\right){\rm S2}$\\
$\vphantom{\bigg|}+\left(-\frac{23}{192}-\frac{65D_s}{768}\right)\zeta_2+\left(-\frac{2173}{768}+\frac{2161D_s}{1152}\right)\zeta_3-\frac{464957}{165888}+\frac{51515D_s}{82944}+\frac{23D_s^2}{1152}\Big]\frac{1}{\e}$
 \end{tabular} \\ \hline
(h) & \begin{tabular}{@{}c@{}} 
$\vspace{-.2cm}\vphantom{\bigg|}-\frac{3}{32}\frac{1}{\e^3}+\left(-\frac{1841}{3072}+\frac{59D_s}{192}-\frac{D_s^2}{48}\right)\frac{1}{\e^2}+\Big[\left(\frac{687}{64}-\frac{21D_s}{4}+\frac{3D_s^2}{8}\right){\rm S2}$ \\
$\vphantom{\bigg|}-\frac{9}{64}\zeta_2+\left(\frac{3347}{2304}-\frac{5D_s}{384}\right)\zeta_3-\frac{144431}{55296}+\frac{13811D_s}{13824}-\frac{17D_s^2}{288}\Big]\frac{1}{\e}$
 \end{tabular} \\ \hline
(i) & \begin{tabular}{@{}c@{}} 
$\vspace{-.2cm}\vphantom{\bigg|}\left(\frac{13}{128}+\frac{D_s}{128}\right)\frac{1}{\e^3}+\left(\frac{4535}{6144}-\frac{265D_s}{768}+\frac{D_s^2}{32}\right)\frac{1}{\e^2}+\Big[\left(-\frac{1779}{128}+\frac{783D_s}{128}-\frac{9D_s^2}{16}\right){\rm S2}$ \\
$\vphantom{\bigg|}+\left(\frac{39}{256}+\frac{3D_s}{256}\right)\zeta_2+\left(-\frac{2263}{2304}+\frac{11D_s}{576}\right)\zeta_3+\frac{311953}{110592}-\frac{7691D_s}{6912}+\frac{5D_s^2}{64}\Big]\frac{1}{\e}$
 \end{tabular} \\ \hline
(j) & \begin{tabular}{@{}c@{}}
 $\vspace{-.2cm}\vphantom{\bigg|}\left(-\frac{3}{32}-\frac{3D_s}{32}\right)\frac{1}{\e^3}+\left(-\frac{41}{32}+\frac{35D_s}{64}\right)\frac{1}{\e^2}+\Big[\left(\frac{927}{32}-\frac{333D_s}{32}\right){\rm S2}$ \\
$ \vphantom{\bigg|}+\left(-\frac{9}{64}-\frac{9D_s}{64}\right)\zeta_2+\left(-\frac{11}{4}+\frac{67D_s}{24}\right)\zeta_3-\frac{1297}{576}+\frac{151D_s}{384}\Big]\frac{1}{\e}$ 
\end{tabular} \\ \hline
(k) & \begin{tabular}{@{}c@{}} 
$\vspace{-.2cm}\vphantom{\bigg|}\left(\frac{1}{64}+\frac{D_s}{64}\right)\frac{1}{\e^3}+\left(\frac{443}{576}-\frac{347D_s}{1152}\right)\frac{1}{\e^2}+\Big[\left(-\frac{985}{64}+\frac{365D_s}{64}\right){\rm S2}$ \\
$ \vphantom{\bigg|}+\left(\frac{3}{128}+\frac{3D_s}{128}\right)\zeta_2+\left(\frac{247}{144}-\frac{13D_s}{12}\right)\zeta_3+\frac{9167}{6912}-\frac{865D_s}{2304}\Big]\frac{1}{\e}$ 
\end{tabular}\\ \hline
(l) & \begin{tabular}{@{}c@{}} 
$\vspace{-.2cm}\vphantom{\bigg|}\left(\frac{5}{64}+\frac{5D_s}{64}\right)\frac{1}{\e^3}+\left(\frac{295}{576}-\frac{283D_s}{1152}\right)\frac{1}{\e^2}+\Big[\left(-\frac{869}{64}+\frac{301D_s}{64}\right){\rm S2}$ \\
$\vphantom{\bigg|}+\left(\frac{15}{128}+\frac{15D_s}{128}\right)\zeta_2+\left(\frac{149}{144}-\frac{41D_s}{24}\right)\zeta_3+\frac{6397}{6912}-\frac{41D_s}{2304}\Big]\frac{1}{\e}$
 \end{tabular} \\
 \hline
 \hline
sum &
$\vphantom{\bigg|} 0 $ \\
\hline
  \end{tabular}
  }
  \end{center}
\caption{ The divergences for the four-graviton amplitude in
  $\NeqFour$ supergravity corresponding to each graph in
  \fig{ThreeLoopFigure}.  To simply the diagrams we chose the
  helicities $(1^-2^-3^+4^+)$ on the pure Yang-Mills side of the
  double-copy decomposition, leaving the states on the
  super-Yang-Mills side arbitrary. On the pure Yang-Mills side, we
  use spinor-helicity with reference momenta
  $q_1=q_2=k_3$ and $q_3=q_4=k_1$.  Each expression includes a
  permutation sum over external legs and the symmetry factor
  appropriate to the graph.  $D_s$ is the state-counting parameter
  and $\eps = (4-D)/2$ is the usual dimensional regularization parameter.
  The transcendental constant $\rm S2$ is defined in \eqn{S2Def}.
  The sum over all contributions in the table vanishes,
  illustrating the phenomenon of enhanced cancellations.
  These results do not include subdivergence subtractions, whose sum
  also vanishes. }
\label{HalfMaxUnsubTable}
\end{table}

\begin{table}[h]
  \begin{center}
  \scalebox{.9}{
\begin{tabular}[t]{||c|c||}
\hline
graph & $\vphantom{\bigg|}(\text{divergence})(4\pi)^6/(\langle 12 \rangle^2 [34]^2 stA^\tree (\frac{\kappa}{2})^8n_{\!f})$ \\
\hline \hline
(a)--(d) & \vphantom{\bigg|}
0 
\\ \hline
(e) & \begin{tabular}{@{}c@{}} 
$\vspace{-.2cm}\vphantom{\bigg|}\left(\frac{43}{192}+\frac{D_s}{32}\right)\frac{1}{\e^3}+\left(\frac{821}{432}-\frac{391D_s}{576}\right)\frac{1}{\e^2}+\Big[\left(-\frac{4627}{192}+\frac{281D_s}{32}\right){\rm S2}$ \\
$\vphantom{\bigg|}+\left(\frac{43}{128}+\frac{3D_s}{64}\right)\zeta_2+\left(-\frac{271}{36}-\frac{17D_s}{12}\right)\zeta_3+\frac{59723}{6912}-\frac{3113D_s}{3456}-\frac{5D_s^2}{16}\Big]\frac{1}{\e}$ 
\end{tabular} \\ \hline
(f) & \begin{tabular}{@{}c@{}} 
$\vspace{-.2cm}\vphantom{\bigg|}\left(\frac{109}{576}-\frac{D_s}{16}\right)\frac{1}{\e^3}+\left(\frac{9}{128}+\frac{D_s}{36}\right)\frac{1}{\e^2}+\Big[\left(-\frac{689}{64}+\frac{17D_s}{8}\right){\rm S2}$ \\
$\vphantom{\bigg|}+\left(\frac{109}{384}-\frac{3D_s}{32}\right)\zeta_2+\left(\frac{425}{144}+\frac{35D_s}{144}\right)\zeta_3-\frac{7649}{5184}-\frac{55D_s}{1728}+\frac{D_s^2}{48}\Big]\frac{1}{\e}$ 
\end{tabular} \\ \hline
(g) & \begin{tabular}{@{}c@{}}
 $\vspace{-.2cm}\vphantom{\bigg|}\left(-\frac{4}{9}+\frac{D_s}{32}\right)\frac{1}{\e^3}+\left(-\frac{7549}{3456}+\frac{43D_s}{64}\right)\frac{1}{\e^2}+\Big[\left(\frac{1849}{48}-\frac{361D_s}{32}\right){\rm S2}$ \\
$\vphantom{\bigg|}+\left(-\frac{2}{3}+\frac{3D_s}{64}\right)\zeta_2+\left(\frac{749}{144}+\frac{163D_s}{144}\right)\zeta_3-\frac{160627}{20736}+\frac{3331D_s}{3456}+\frac{7D_s^2}{24}\Big]\frac{1}{\e}$ 
\end{tabular} \\ \hline
(h) & 
$\vphantom{\bigg|}\left(-\frac{15}{128}+\frac{D_s}{24}\right)\frac{1}{\e^2}+\Big[\left(\frac{9}{8}-\frac{3D_s}{4}\right){\rm S2}+\left(\frac{1}{24}+\frac{95D_s}{144}\right)\zeta_3+\frac{3481}{6912}-\frac{599D_s}{1728}+\frac{D_s^2}{16}\Big]\frac{1}{\e}$
 \\ \hline
(i) &
 $\vphantom{\bigg|}\frac{1}{32}\frac{1}{\e^3}+\left(\frac{127}{384}-\frac{D_s}{16}\right)\frac{1}{\e^2}+\Big[\left(-\frac{153}{32}+\frac{9D_s}{8}\right){\rm S2}+\frac{3}{64}\zeta_2+\left(-\frac{2}{3}-\frac{89D_s}{144}\right)\zeta_3+\frac{179}{2304}+\frac{545D_s}{1728}-\frac{D_s^2}{16}\Big]\frac{1}{\e}$
 \\ \hline
(j) & 
$\vphantom{\bigg|}-\frac{3}{8}\frac{1}{\e^3}+\left(-\frac{97}{48}+\frac{17D_s}{24}\right)\frac{1}{\e^2}+\Big[\left(\frac{255}{8}-\frac{21D_s}{2}\right){\rm S2}-\frac{9}{16}\zeta_2+\left(\frac{43}{6}+\frac{5D_s}{3}\right)\zeta_3-\frac{757}{96}+\frac{151D_s}{144}+\frac{D_s^2}{3}\Big]\frac{1}{\e}$
 \\ \hline
(k) & 
$\vphantom{\bigg|}\frac{1}{16}\frac{1}{\e^3}+\left(\frac{337}{288}-\frac{3D_s}{8}\right)\frac{1}{\e^2}+\Big[\left(-\frac{265}{16}+\frac{45D_s}{8}\right){\rm S2}+\frac{3}{32}\zeta_2+\left(-\frac{13}{6}-\frac{5D_s}{6}\right)\zeta_3+\frac{707}{192}-\frac{23D_s}{48}-\frac{D_s^2}{6}\Big]\frac{1}{\e}$ 
\\ \hline
(l) &
 $\vphantom{\bigg|}\frac{5}{16}\frac{1}{\e^3}+\left(\frac{245}{288}-\frac{D_s}{3}\right)\frac{1}{\e^2}+\Big[\left(-\frac{245}{16}+\frac{39D_s}{8}\right){\rm S2}+\frac{15}{32}\zeta_2+\left(-5-\frac{5D_s}{6}\right)\zeta_3+\frac{269}{64}-\frac{41D_s}{72}-\frac{D_s^2}{6}\Big]\frac{1}{\e}$
 \\ \hline
 \hline
sum  &
$\vphantom{\bigg|} 0 $ \\
\hline
  \end{tabular}
  }
  \end{center}
\caption{Additional diagrammatic contributions appearing in the
  four-graviton amplitude of $\NeqFive$ supergravity.  This
  contribution contains also the $n_{\!f}$ state-counting
  parameter. The total $\NeqFive$ divergence is given by the sum over
  these contributions and those in \tab{HalfMaxUnsubTable}.  The
  vanishing of the sum over the entries in each table individually is
  a reflection of enhanced cancellations.  Subdivergences
  automatically cancel amongst themselves and are not included.  The
  choice of external helicity states and reference momenta are as in
  \tab{HalfMaxUnsubTable}.}
\label{FermUnsubTable}
\end{table}

As a warm up to our four-loop calculation, we first present the
corresponding three-loop calculation in $\NeqFive$ supergravity.  We
follow the same techniques summarized in Sect.~\ref{MethodsSection}
and described in some detail in Ref.~\cite{N4Mat}.  In contrast to
$\NeqFour$ supergravity, in this case we should be able to construct a
covariant integrand that is manifestly ultraviolet finite, bypassing
the need for loop integration to demonstrate that it is ultraviolet
finite.  However, we do not do so here.  Instead, we proceed the same
way as at four loops by first computing the $\NeqFour$ supergravity
divergences and then adding in the extra contributions needed in
$\NeqFive$ supergravity.  This allows us to observe enhanced
cancellations.  In fact, we are able to show finiteness with the
enhanced cancellations alone, even without accounting for
cancellations arising from the extra supersymmetry in the $\NeqFive$
theory compared to the $\NeqFour$ theory.  Thus, the cancellations are
stronger than those required to demonstrate finiteness.

In the calculation, we leave two state-counting parameters to make it
simple to switch between various supergravity theories.  The first
parameter is $D_s$, which is obtained from contractions of the metric
$\eta_{\mu \nu}$ from the Lorentz algebra, while the second is $n_{\!
  f}$, which counts the number of Majorana fermions added to the pure
Yang-Mills side of the double copy.  By choosing $D_s = 4$ and $n_{\!
  f} =0$, we obtain pure $\NeqFour$ supergravity.  By setting the
parameters to $D_s = 4$ and $n_{\!  f} =1$, we obtain $\NeqFive$
supergravity.  We can also obtain results for $\NeqFour$ supergravity
with $n_\V$ matter multiplets by choosing $D_s = 4+n_\V$ and $n_{\! f}
=0$, where $n_\V$ is the number of internal matter vector
multiplets~\cite{N4Mat}.

Our $D=4$ divergence-calculation results are summarized in
\tabs{HalfMaxUnsubTable}{FermUnsubTable} with the results
corresponding to each graph in Fig.~\ref{ThreeLoopFigure} used to
organize these calculations.  The 12 diagrams correspond to the
nonvanishing ones of $\NeqFour$ super-Yang-Mills theory in the BCJ
representation~\cite{BCJLoop}.  \Tab{HalfMaxUnsubTable} contains all
contributions that do not depend on the parameter $n_{\!  f}$, and
\tab{FermUnsubTable} contains all the pieces that do depend on $n_{\!
  f}$.  In the calculation, we take the external states to be gluons
on the $\NeqZero$ or $\NeqOne$ side of the double copy, keeping the
polarizations vectors formal.  However, to simplify the tables we
apply four-dimensional spinor-helicity (see Ref.~\cite{SpinorHelicity}
for a recent review) on the polarization vectors and specify the
external-gluon states to be ${-}{-}{+}{+}$.  We also make convenient
choices of reference momenta: $q_1=q_2=k_3$ and $q_3=q_4=k_1$.  This
choice makes the divergences in diagrams (a)--(d) vanish for both
tables.  It also results in terms containing a factor of $n_{\!f}^2$,
from two fermion loops, to vanish individually in all diagrams in
Table~\ref{FermUnsubTable} (instead of in the sum over diagrams).  We
have not included subdivergence subtractions in the tables, but we
have explicitly confirmed that, with the use of the uniform mass
regulator, the subdivergences cancel as expected, given that there are
no lower-loop divergences.  In the tables, the $\zeta_i$ are the
standard Riemann zeta constants. The transcendental constant ${\rm
  S2}$ appearing in the tables is
\begin{equation}
{\rm S2} = \frac{4}{9 \sqrt{3}} {\rm Cl}_2 \left(\frac{\pi}{3} \right)\,,
\label{S2Def}
\end{equation}
where ${\rm Cl}_2(x) = {\rm Im}({\rm Li}_2(e^{ix}))$ is the Clausen
function.  As the tables illustrate, when one sums over all diagrams,
the result is finite for any choice of the state-counting parameters.

Although we made special helicity choices for the tables, our
calculation is based on using formal polarization states and is
therefore valid for any external state that is a direct
product of a gluon state and an $\NeqFour$ super-Yang-Mills state.
This corresponds to a subset of the $\NeqFive$ supergravity states.
Nevertheless, the result also extends to {\it any} $\NeqFive$ state
because the $\NeqOne$ super-Yang-Mills supersymmetry
identities~\cite{SWI} are powerful enough to relate all
four-point amplitudes to the gluonic ones.  (A discussion of these
identities at two loops is given in Ref.~\cite{TwoLoopGluino}.)  It is
interesting that for {\it any} values of the state-counting
parameters, the divergences vanish.

Summarizing, not only is $\NeqFour$ supergravity ultraviolet finite at
three loops, but the extra pieces needed to obtain $\NeqFive$
supergravity using the decomposition (\ref{NeqOneSusyDecomp}) are
finite by themselves:
\begin{align}
\mathcal{M}_4^{\threeloop}\big|_{\mathcal{N}=4,\,\mathrm{div.}}&=0\,, 
\notag \\
\mathcal{M}_4^{\threeloop}\big|_{(\mathcal{N}=5\, - \,
                                {\mathcal{N}=4),\,\mathrm{div.}}} &=0\,.
\label{enhancedThreeLoop}
\end{align}
The independent vanishings in \eqn{enhancedThreeLoop} show that the
extra supersymmetry of the $\NeqFive$ theory compared to $\NeqFour$
theory is not needed to make $\NeqFive$ supergravity finite.  While
this is no surprise given the three-loop finiteness of $\NeqFour$
supergravity, it does explicitly demonstrate that ultraviolet cancellations
exist in subpieces for which there is no power-counting argument.   Thus, the
$\NeqFive$ case is another explicit example of enhanced ultraviolet
cancellations that go beyond the ones that have been understood by
any standard-symmetry considerations.

\section{Four loops}
\label{FourLoopSection}

We now consider four loops. We first summarize the calculation of the
$\NeqFour$ supergravity four-point divergence presented in
Ref.~\cite{FourLoopN4}, giving a few additional intermediate results.  We
then turn to the corresponding calculation in $\NeqFive$ supergravity,
showing that the divergence vanishes.

\begin{figure}[tbh]
\centering
\renewcommand{\thesubfigure}{(\arabic{subfigure})}
\renewcommand{\subfigcapskip}{-.1cm}
\renewcommand{\subfigbottomskip}{-.1cm}
\begin{align}
&
\setcounter{subfigure}{0}
\subfigure[]{\hspace{-0.1cm}\includegraphics[clip,scale=0.32]{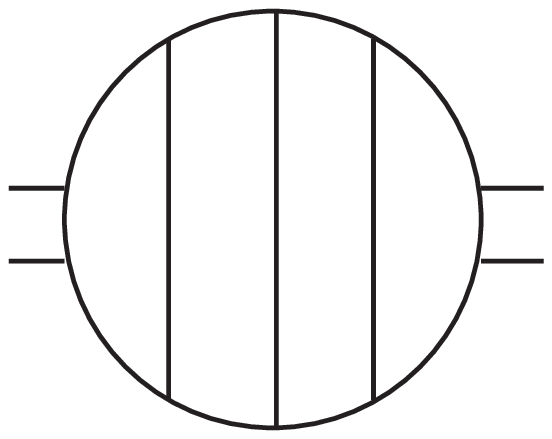}}
\hspace{0.4 cm}
\subfigure[]{\hspace{-0.1cm}\includegraphics[clip,scale=0.32]{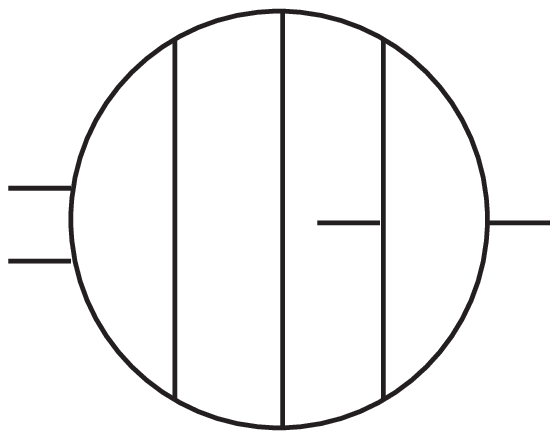}}
\hspace{0.4cm}
\subfigure[]{\hspace{-0.15cm}\includegraphics[clip,scale=0.32]{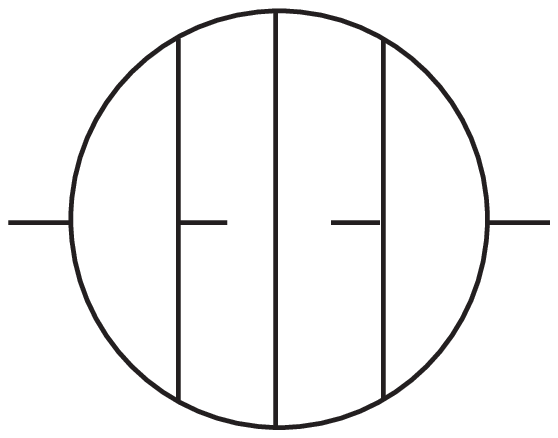}}
\hspace{0.65cm}
\subfigure[]{\hspace{-0.25cm}\includegraphics[clip,scale=0.32]{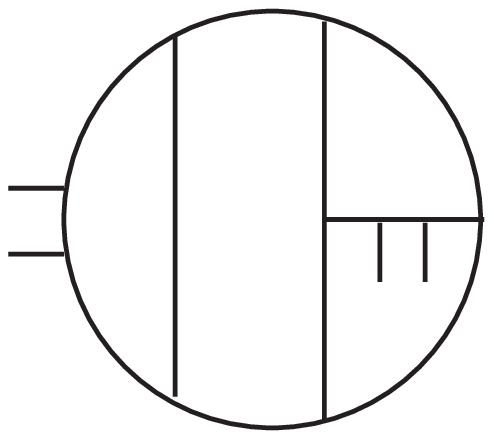}}
\hspace{0.7cm}
\subfigure[]{\hspace{-0.3cm}\includegraphics[clip,scale=0.32]{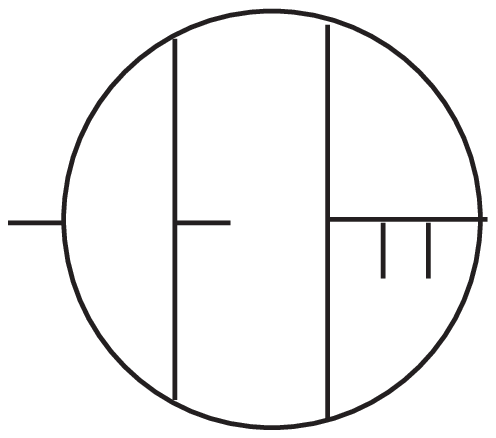}}
\hspace{0.8cm}
\subfigure[]{\hspace{-0.3cm}\includegraphics[clip,scale=0.32]{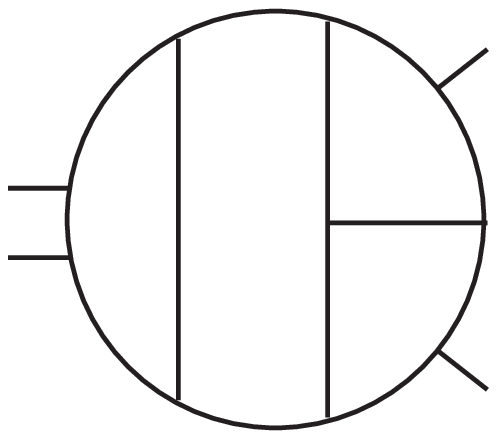}}
\nonumber\\
&\hskip .1 cm 
\subfigure[]{\hspace{-0.2cm}\includegraphics[clip,scale=0.32]{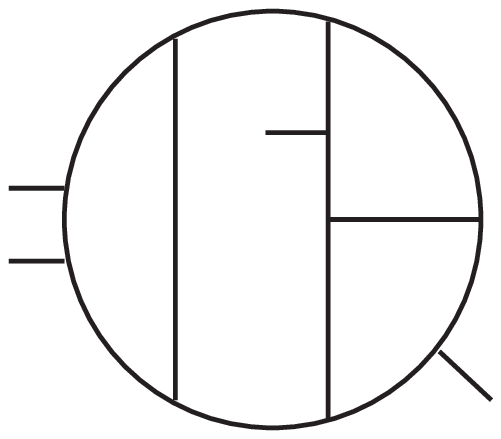}}
\hspace{0.8cm}
\subfigure[]{\hspace{-0.25cm}\includegraphics[clip,scale=0.32]{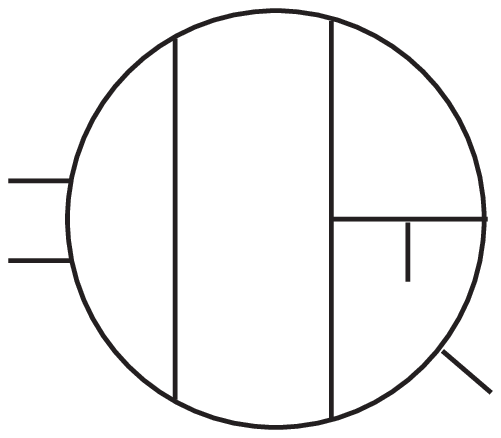}}
\hspace{0.8cm}
\subfigure[]{\hspace{-0.25cm}\includegraphics[clip,scale=0.32]{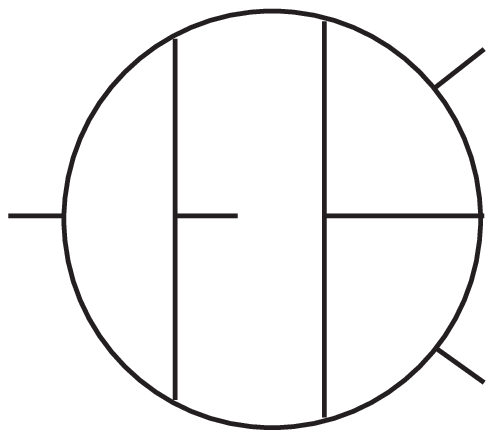}}
\hspace{0.7cm}
\subfigure[]{\hspace{-0.25cm}\includegraphics[clip,scale=0.32]{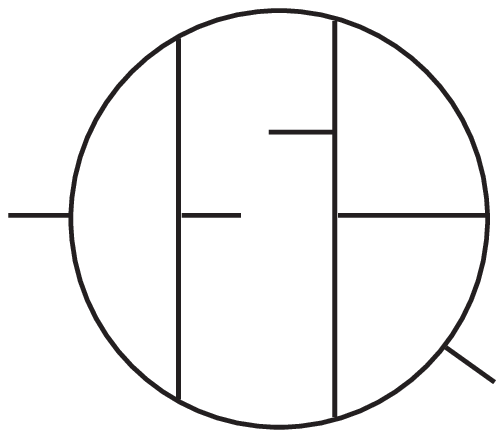}}
\hspace{0.6cm}
\subfigure[]{\hspace{-0.25cm}\includegraphics[clip,scale=0.32]{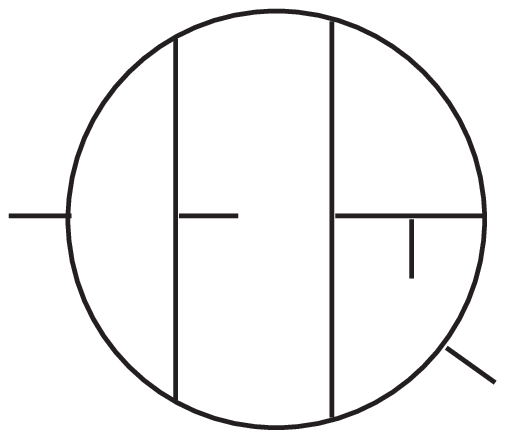}}
\hspace{0.6cm}
\subfigure[]{\hspace{-0.1cm}\includegraphics[clip,scale=0.32]{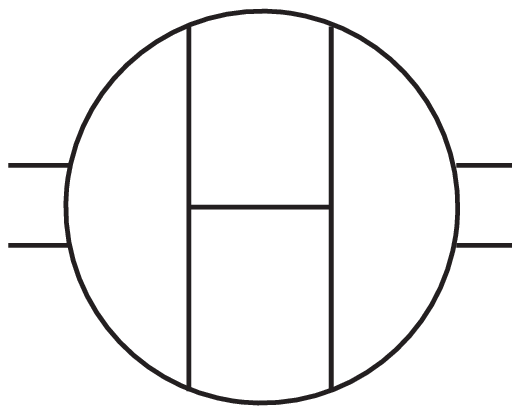}}
\nonumber\\
&
\hspace{0.2cm}
\subfigure[]{\hspace{-0.3cm}\includegraphics[clip,scale=0.32]{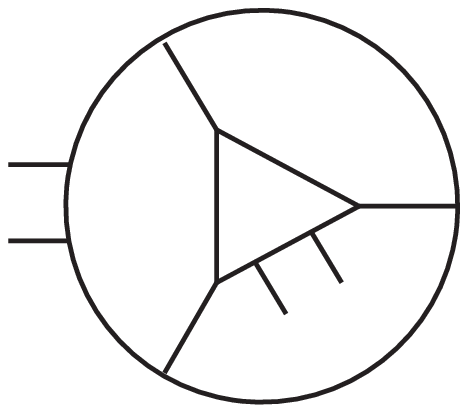}}
\hspace{0.8cm}
\subfigure[]{\hspace{-0.15cm}\includegraphics[clip,scale=0.32]{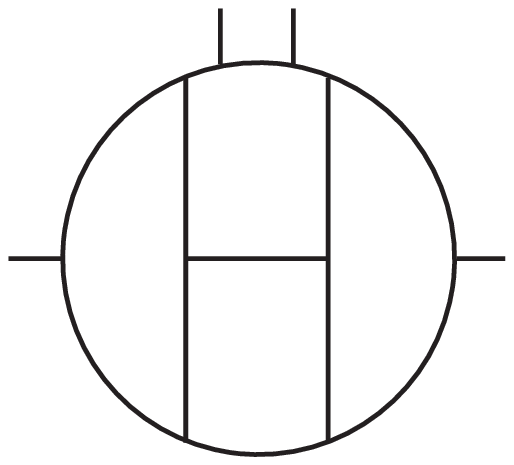}}
\hspace{0.8cm}
\subfigure[]{\hspace{-0.3cm}\includegraphics[clip,scale=0.32]{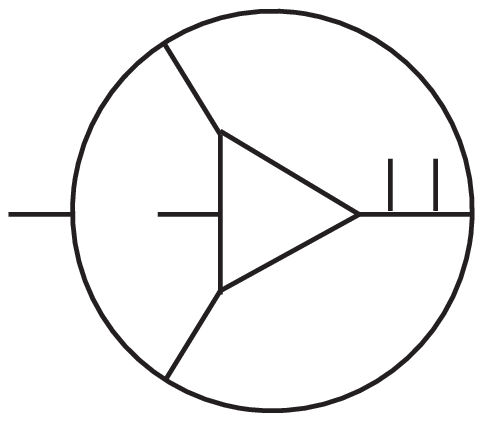}}
\hspace{0.8cm}
\subfigure[]{\hspace{-0.3cm}\includegraphics[clip,scale=0.32]{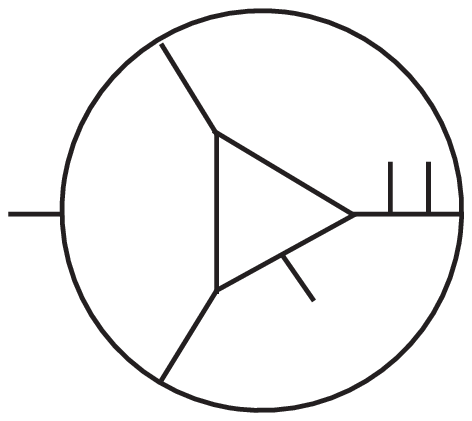}}
\hspace{0.8cm}
\subfigure[]{\hspace{-0.3cm}\includegraphics[clip,scale=0.32]{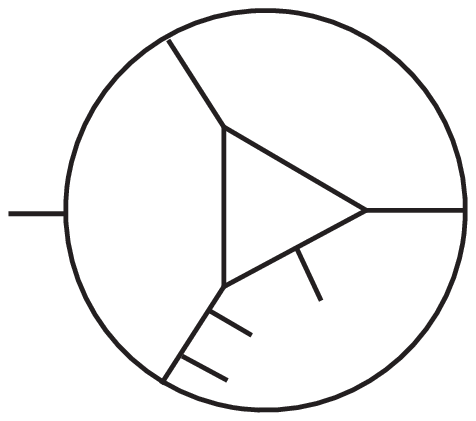}}
\hspace{0.7cm}
\subfigure[]{\hspace{-0.2cm}\includegraphics[clip,scale=0.32]{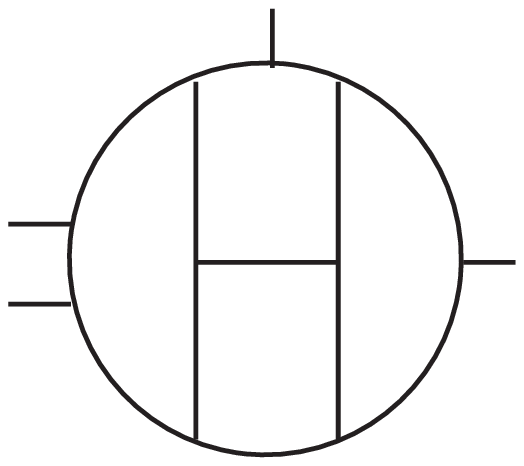}}
\nonumber\\
&
\hspace{0.1cm}
\subfigure[]{\hspace{-0.1cm}\includegraphics[clip,scale=0.32]{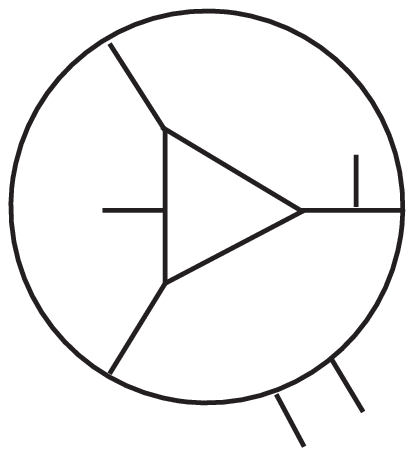}}
\hspace{1.cm}
\subfigure[]{\hspace{-0.2cm}\includegraphics[clip,scale=0.32]{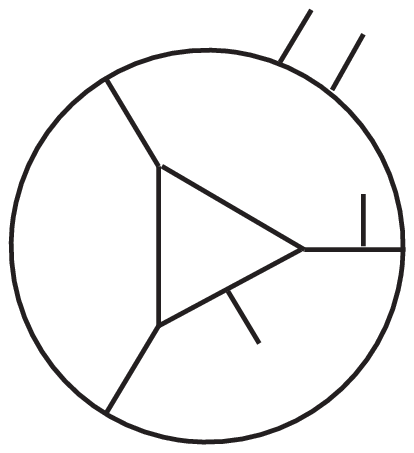}}
\hspace{1.1cm}
\subfigure[]{\hspace{-0.15cm}\includegraphics[clip,scale=0.32]{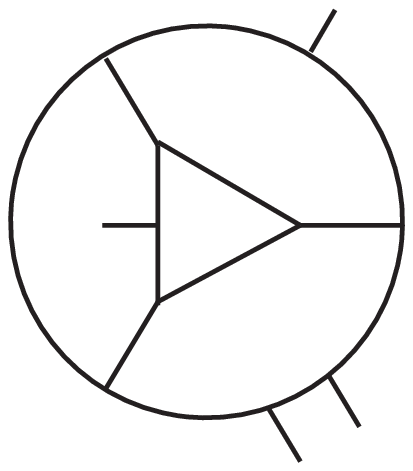}}
\hspace{0.8cm}
\subfigure[]{\hspace{-0.2cm}\includegraphics[clip,scale=0.32]{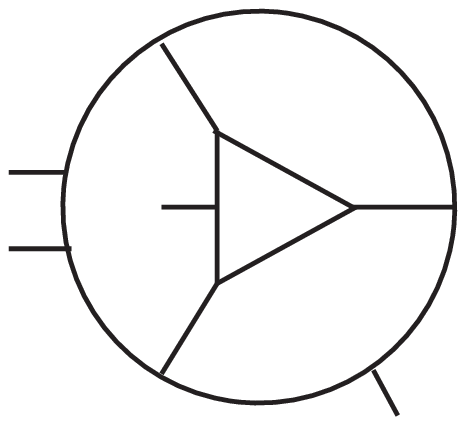}}
\hspace{0.8cm}
\subfigure[]{\hspace{-0.2cm}\includegraphics[clip,scale=0.32]{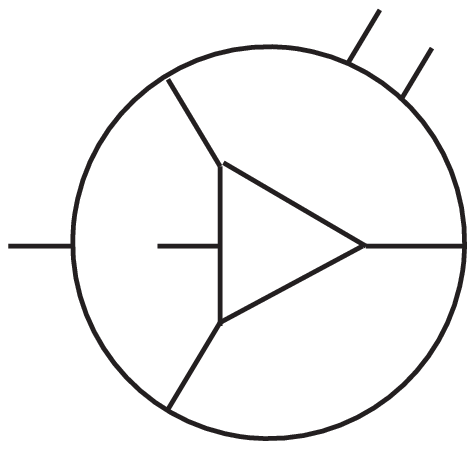}}
\hspace{0.7cm}
\subfigure[]{\hspace{-0.2cm}\includegraphics[clip,scale=0.32]{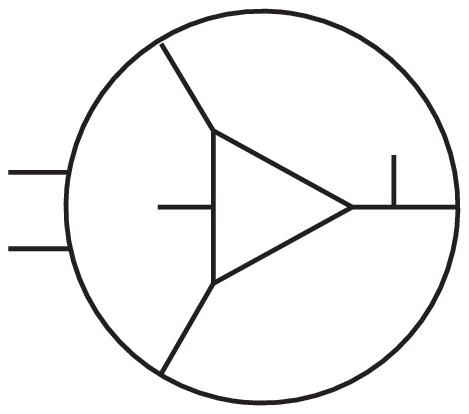}}
\nonumber\\
&
\hspace{0.1cm}
\subfigure[]{\hspace{-0.2cm}\includegraphics[clip,scale=0.32]{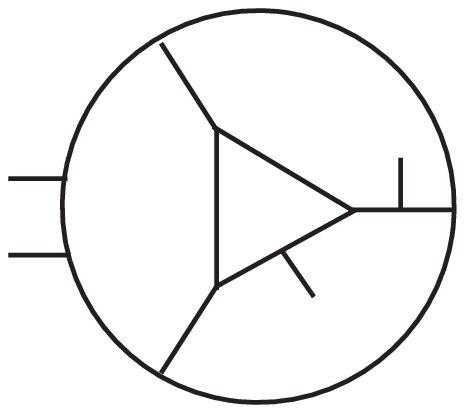}}
\hspace{0.8cm}
\subfigure[]{\hspace{-0.1cm}\includegraphics[clip,scale=0.32]{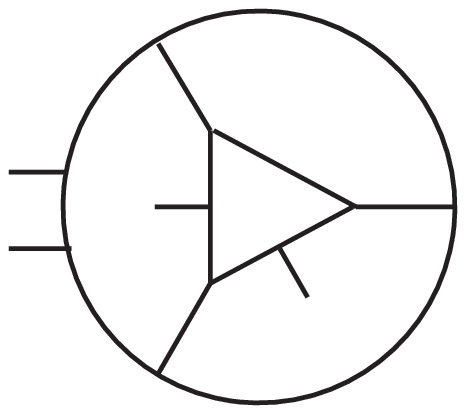}}
\hspace{0.8 cm}
\subfigure[]{\hspace{-0.1cm}\includegraphics[clip,scale=0.32]{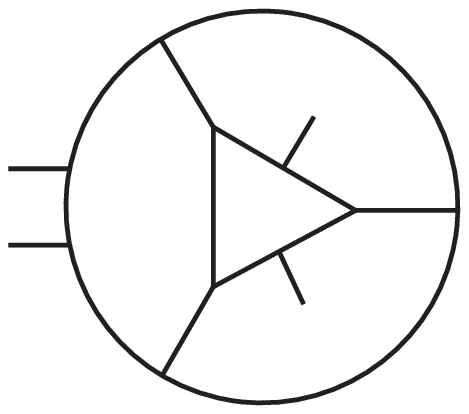}}
\hspace{0.6cm}
\subfigure[]{\hspace{-0.1cm}\includegraphics[clip,scale=0.32]{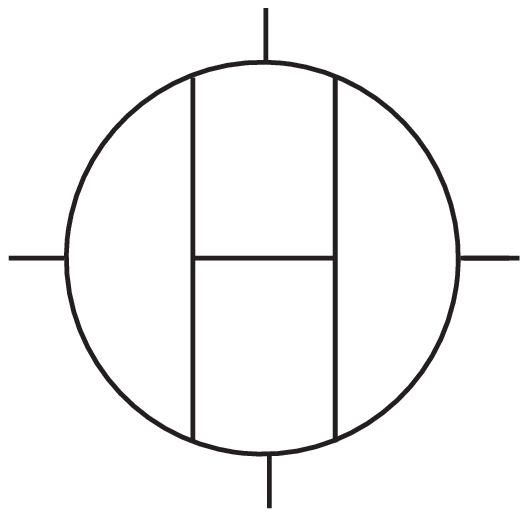}}
\hspace{0.7cm}
\subfigure[]{\hspace{-0.2cm}\includegraphics[clip,scale=0.32]{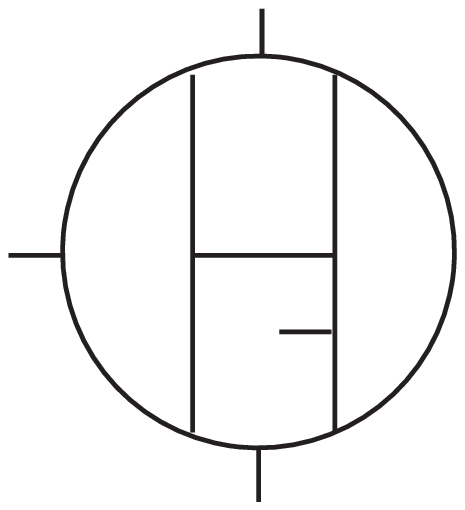}}
\hspace{0.6cm}
\subfigure[]{\hspace{-0.2cm}\includegraphics[clip,scale=0.32]{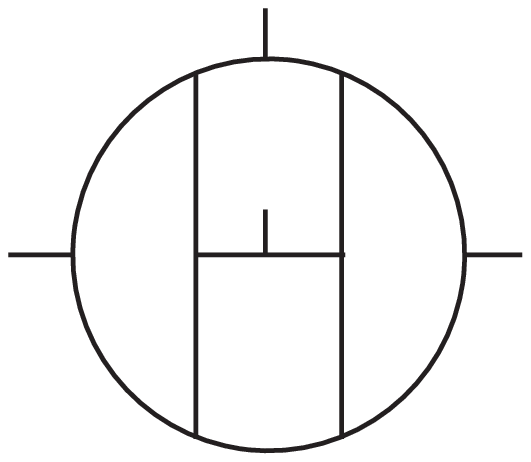}}
\nonumber\\
&
\hspace{0.1cm}
\subfigure[]{\hspace{-0.1cm}\includegraphics[clip,scale=0.32]{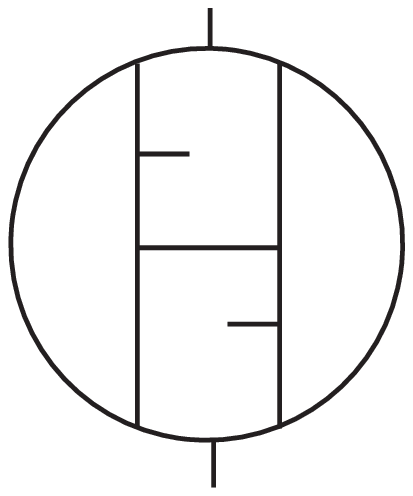}}
\hspace{0.9cm}
\subfigure[]{\hspace{-0.1cm}\includegraphics[clip,scale=0.32]{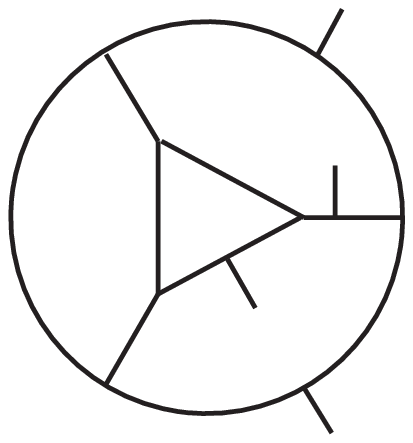}}
\hspace{1.cm}
\subfigure[]{\hspace{-0.3cm}\includegraphics[clip,scale=0.32]{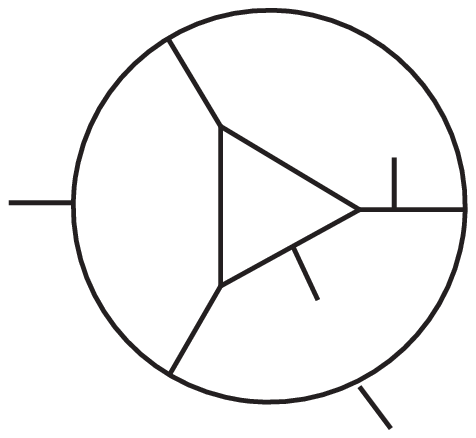}}
\hspace{1.0cm}
\subfigure[]{\hspace{-0.1cm}\includegraphics[clip,scale=0.32]{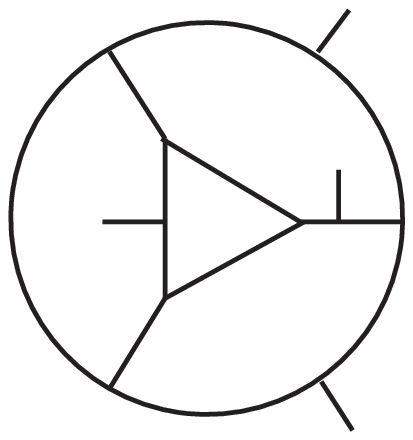}}
\hspace{0.8cm}
\subfigure[]{\hspace{-0.3cm}\includegraphics[clip,scale=0.32]{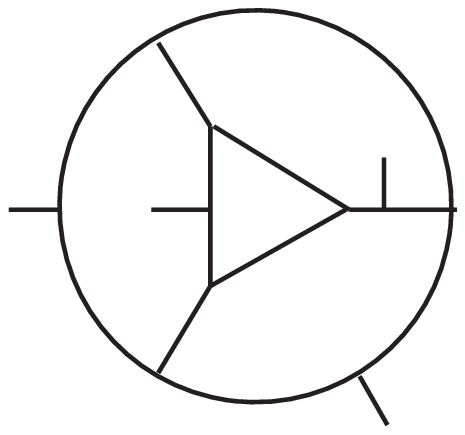}}
\hspace{0.8cm}
\subfigure[]{\hspace{-0.3cm}\includegraphics[clip,scale=0.32]{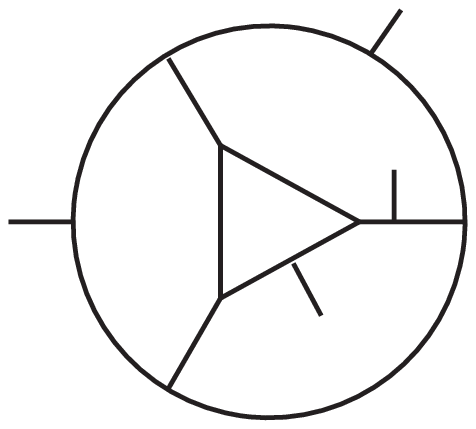}}
\nonumber \\
&
\hspace{0.1cm}
\subfigure[]{\hspace{-0.2cm}\includegraphics[clip,scale=0.32]{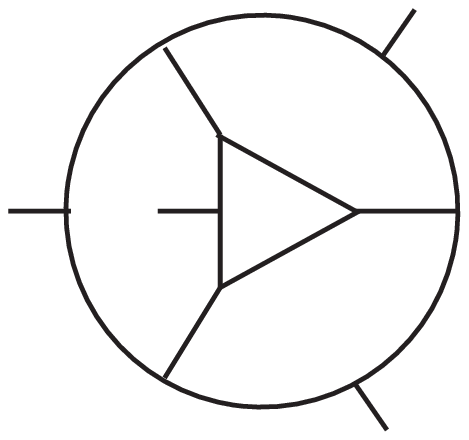}}
\hspace{0.8cm}
\subfigure[]{\hspace{-0.2cm}\includegraphics[clip,scale=0.32]{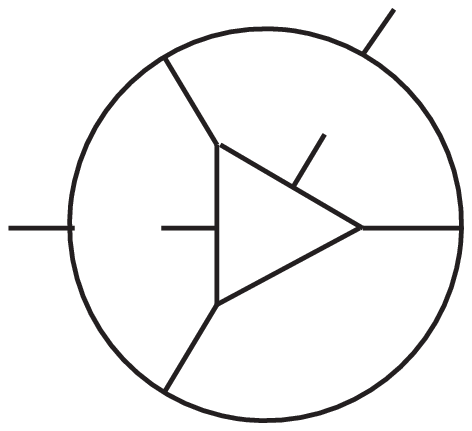}}
\hspace{0.8cm}
\subfigure[]{\hspace{-0.2cm}\includegraphics[clip,scale=0.31]{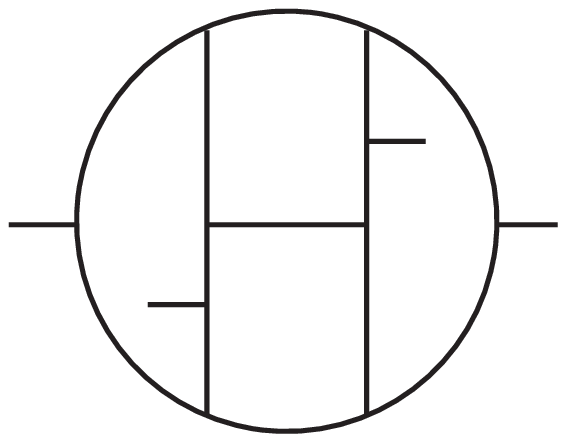}}
\hspace{0.7cm}
\subfigure[]{\hspace{-0.2cm}\includegraphics[clip,scale=0.30]{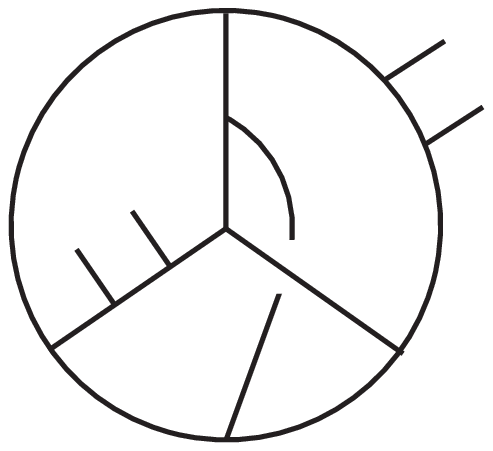}}
\hspace{0.9cm}
\subfigure[]{\hspace{-0.2cm}\includegraphics[clip,scale=0.30]{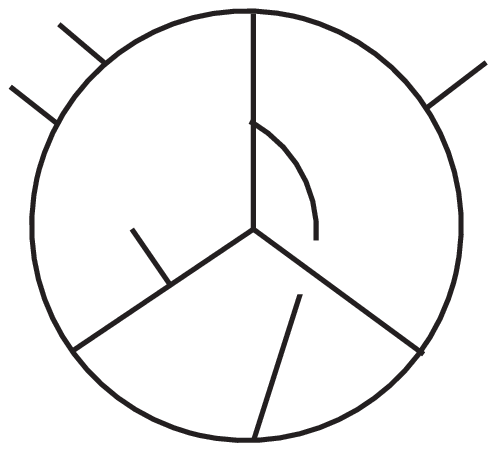}}
\hspace{0.7cm}
\subfigure[]{\hspace{-0.2cm}\includegraphics[clip,scale=0.30]{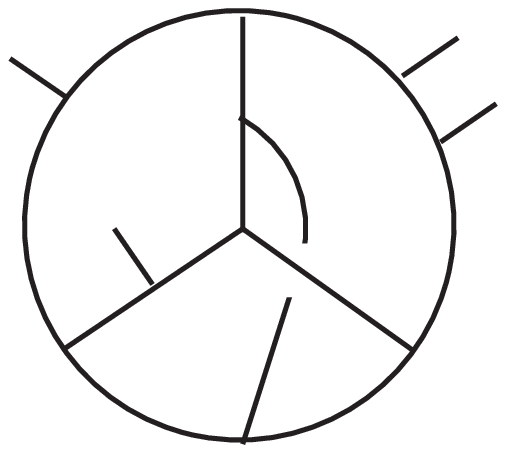}}
\nonumber
\end{align}
\caption[a]{The first 42 diagrams for the four-loop four-point
  amplitudes of $\NeqFour$ and $\NeqFive$ supergravity. These correspond to the
  $\NeqFour$ super-Yang-Mills diagrams of Ref.~\cite{ck4l}. }
\label{FourLoop1Figure}
\end{figure}

\begin{figure}[tbh]
\centering
\renewcommand{\thesubfigure}{(\arabic{subfigure})}
\renewcommand{\subfigcapskip}{-.1cm}
\renewcommand{\subfigbottomskip}{-.1cm}
\begin{align}
&
\setcounter{subfigure}{42}
\hspace{0.2cm}
\subfigure[]{\hspace{-0.1cm}\includegraphics[clip,scale=0.30]{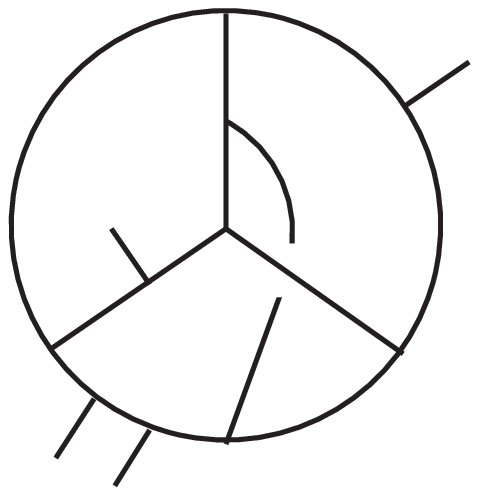}}
\hspace{0.8cm}\vspace{10 cm}
\subfigure[]{\hspace{-0.1cm}\includegraphics[clip,scale=0.30]{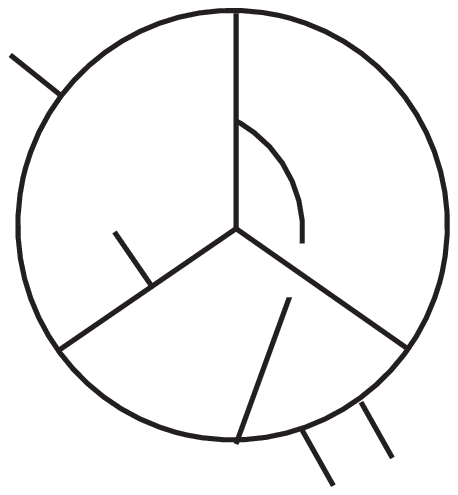}}
\hspace{0.8cm}
\subfigure[]{\hspace{-0.1cm}\includegraphics[clip,scale=0.31]{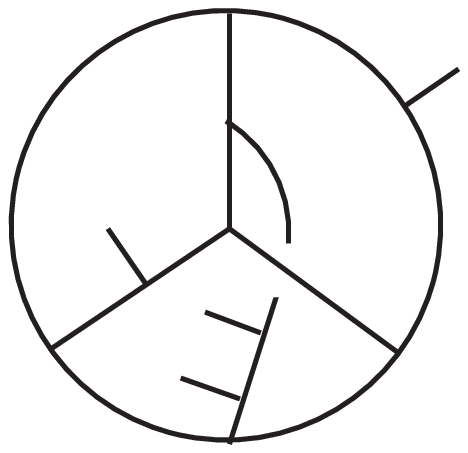}}
\hspace{0.7cm}
\subfigure[]{\hspace{-0.1cm}\includegraphics[clip,scale=0.30]{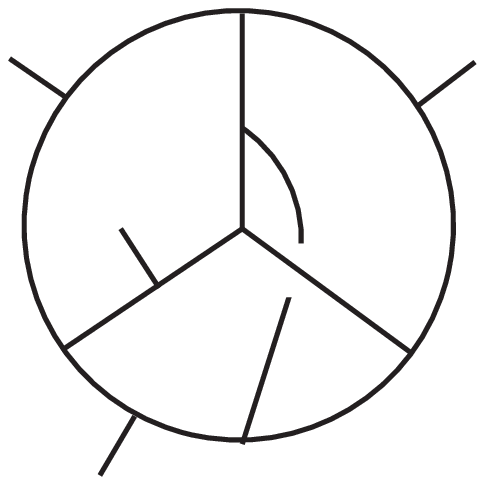}}
\hspace{0.8cm}
\subfigure[]{\hspace{-0.1cm}\includegraphics[clip,scale=0.30]{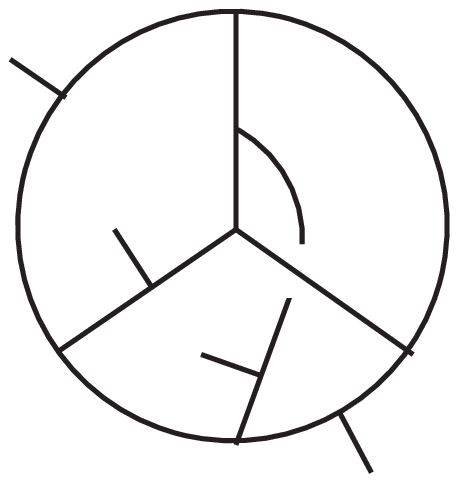}}
\hspace{0.7cm}
\subfigure[]{\hspace{-0.1cm}\includegraphics[clip,scale=0.30]{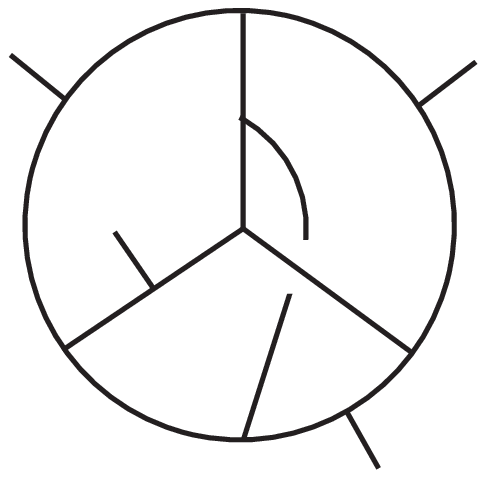}}
\nonumber\\
&
\hspace{0.2cm}
\subfigure[]{\hspace{-0.1cm}\includegraphics[clip,scale=0.30]{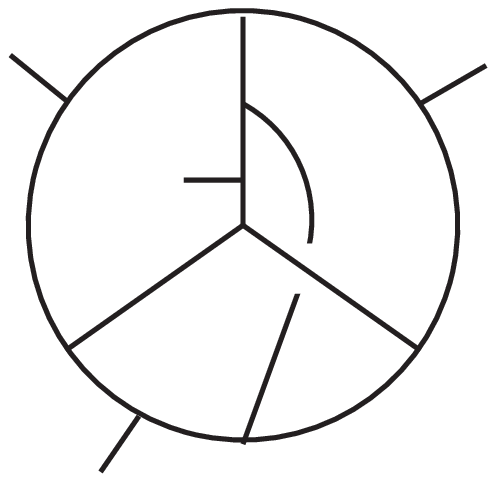}}
\hspace{0.7cm}
\subfigure[]{\hspace{-0.1cm}\includegraphics[clip,scale=0.30]{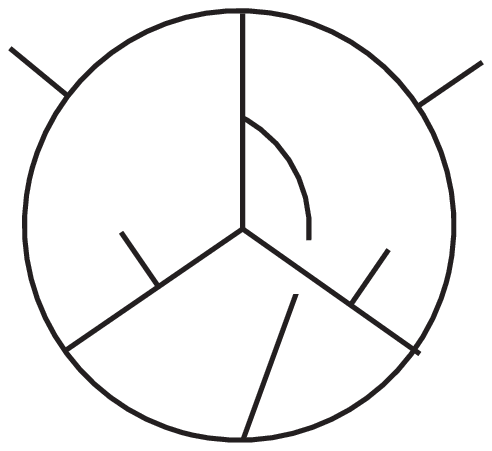}}
\hspace{0.6cm}
\subfigure[]{\hspace{-0.1cm}\includegraphics[clip,scale=0.33]{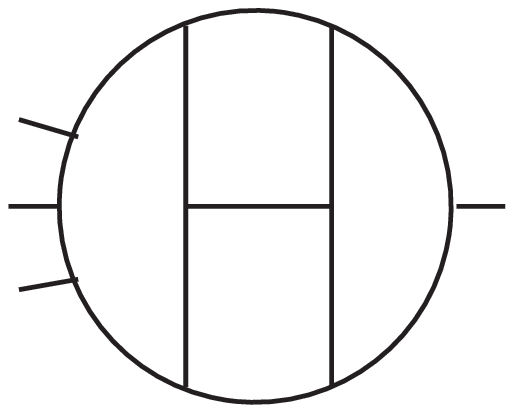}}
\hspace{0.6 cm}
\subfigure[]{\hspace{-0.2cm}\includegraphics[clip,scale=0.32]{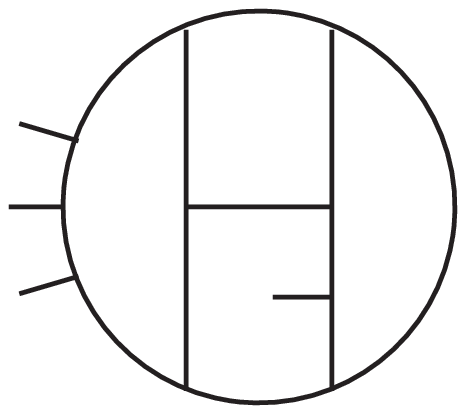}}
\hspace{0.7cm}
\subfigure[]{\hspace{-0.3cm}\includegraphics[clip,scale=0.32]{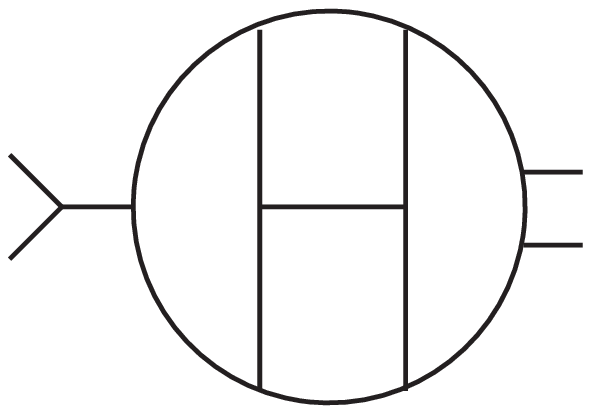}}
\hspace{0.5cm}
\subfigure[]{\hspace{-0.3cm}\includegraphics[clip,scale=0.32]{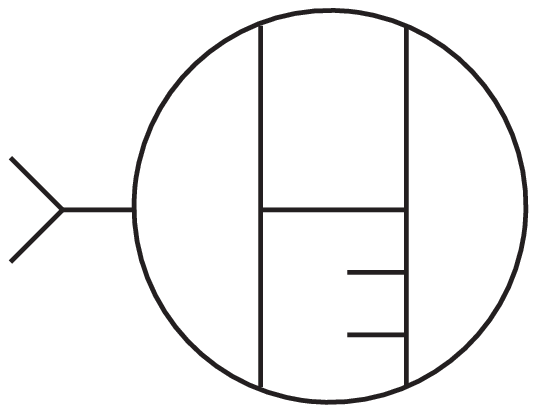}}
\nonumber\\
&
\hspace{0.15cm}
\subfigure[]{\hspace{-0.3cm}\includegraphics[clip,scale=0.32]{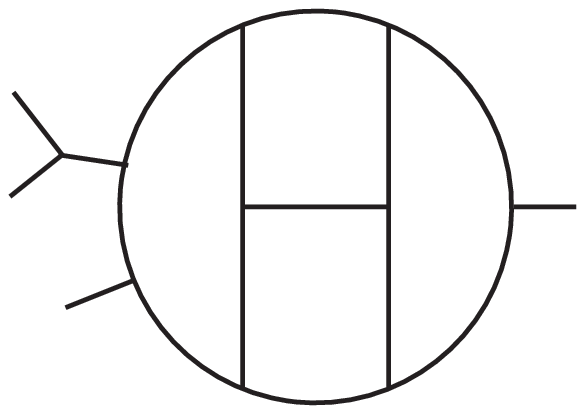}}
\hspace{0.55cm}
\subfigure[]{\hspace{-0.3cm}\includegraphics[clip,scale=0.32]{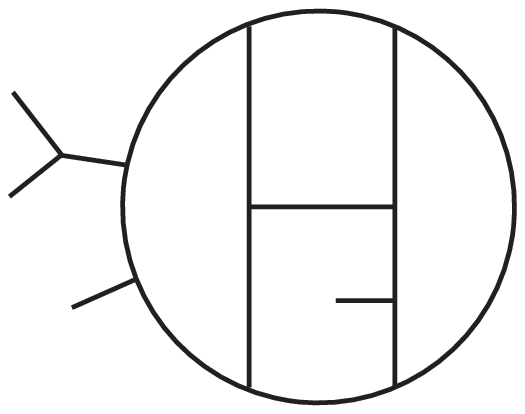}}
\hspace{0.7cm}
\subfigure[]{\hspace{-0.3cm}\includegraphics[clip,scale=0.32]{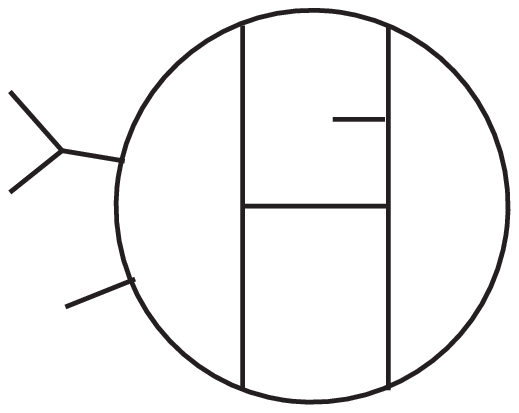}}
\hspace{0.6cm}
\subfigure[]{\hspace{-0.3cm}\includegraphics[clip,scale=0.32]{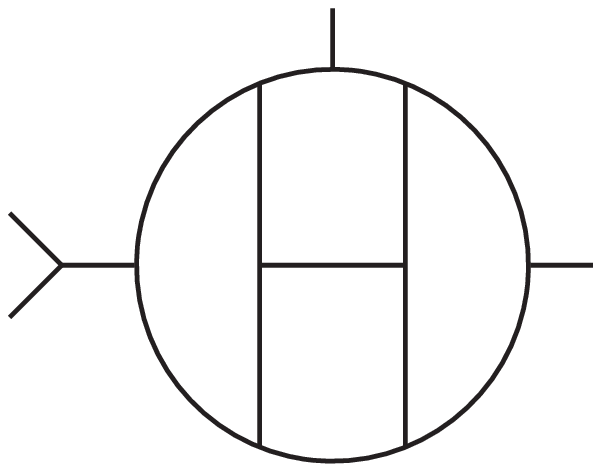}}
\hspace{0.6cm}
\subfigure[]{\hspace{-0.3cm}\includegraphics[clip,scale=0.32]{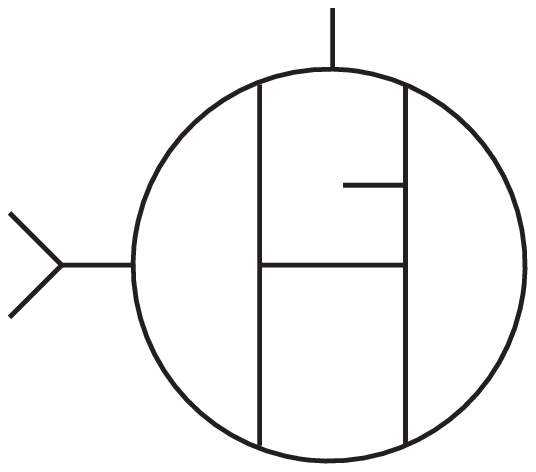}}
\hspace{0.6cm}
\subfigure[]{\hspace{-0.3cm}\includegraphics[clip,scale=0.32]{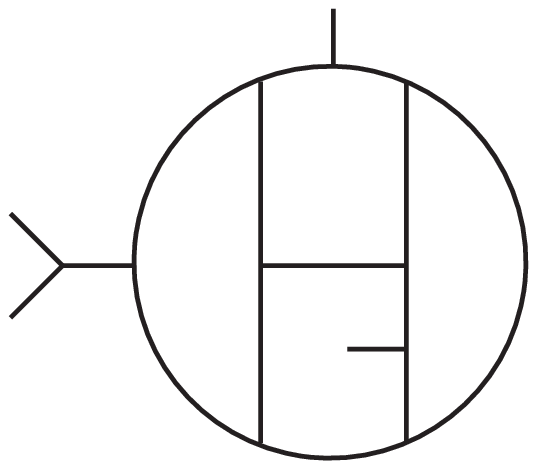}}
\nonumber\\
&
\hspace{0.15cm}
\subfigure[]{\hspace{-0.3cm}\includegraphics[clip,scale=0.32]{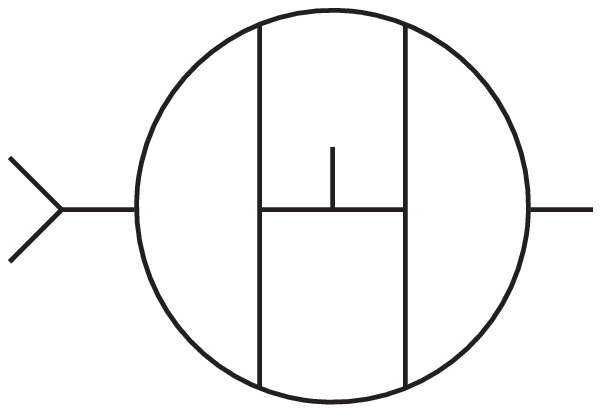}}
\hspace{0.5cm}
\subfigure[]{\hspace{-0.3cm}\includegraphics[clip,scale=0.32]{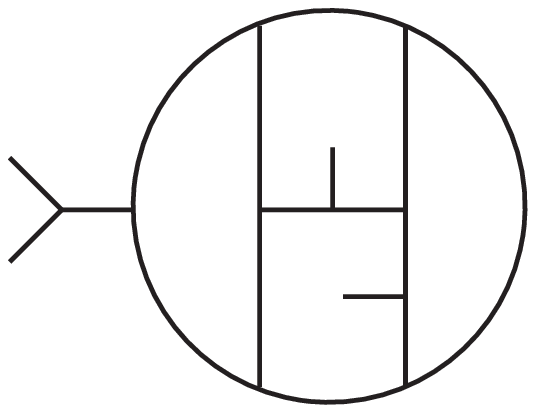}}
\hspace{0.7cm}
\subfigure[]{\hspace{-0.3cm}\includegraphics[clip,scale=0.32]{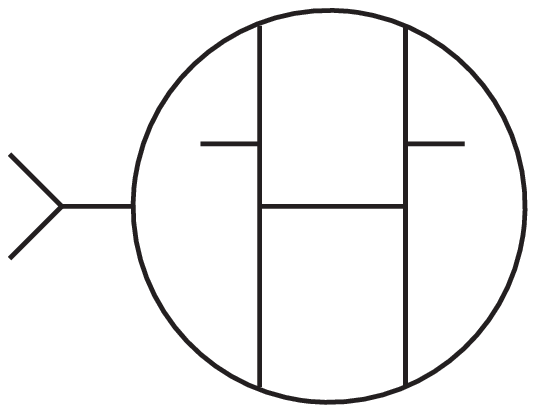}}
\hspace{0.6cm}
\subfigure[]{\hspace{-0.3cm}\includegraphics[clip,scale=0.32]{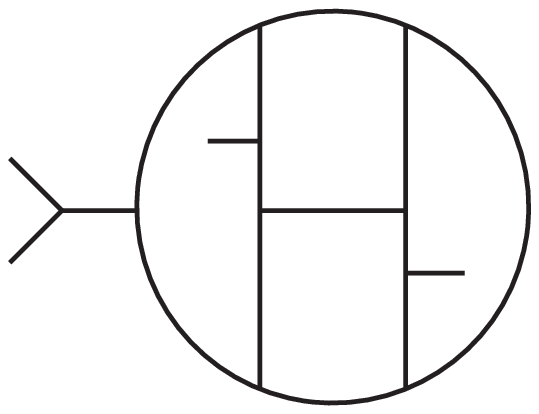}}
\hspace{0.6cm}
\subfigure[]{\hspace{-0.3cm}\includegraphics[clip,scale=0.32]{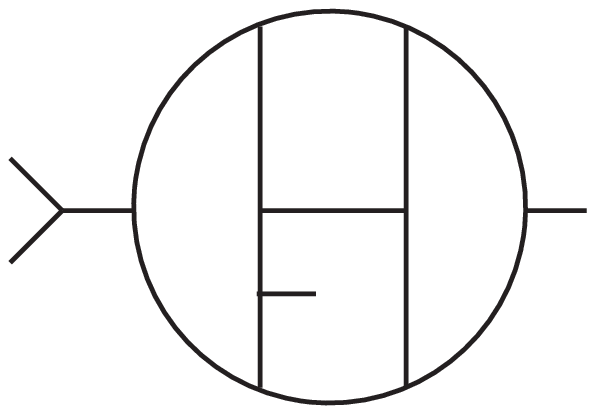}}
\hspace{0.55cm}
\subfigure[]{\hspace{-0.3cm}\includegraphics[clip,scale=0.32]{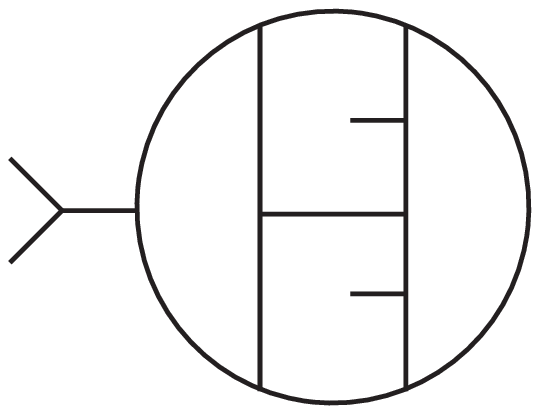}}
\nonumber\\
&
\hspace{0.2cm}
\subfigure[]{\hspace{-0.3cm}\includegraphics[clip,scale=0.32]{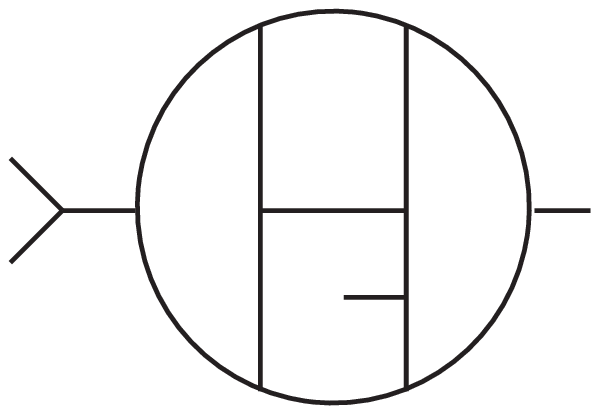}}
\hspace{0.55cm}
\subfigure[]{\hspace{-0.3cm}\includegraphics[clip,scale=0.32]{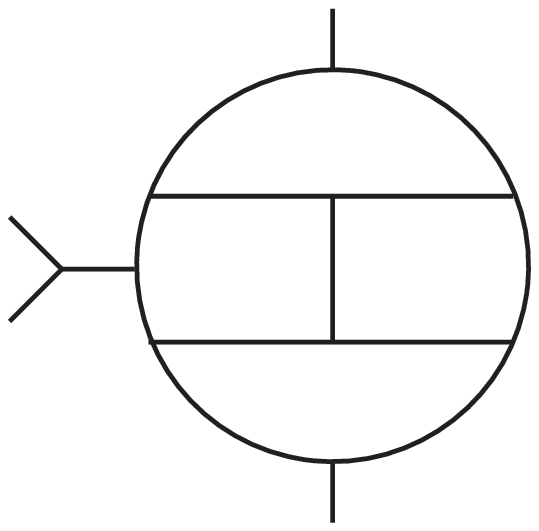}}
\hspace{0.6cm}
\subfigure[]{\hspace{-0.3cm}\includegraphics[clip,scale=0.32]{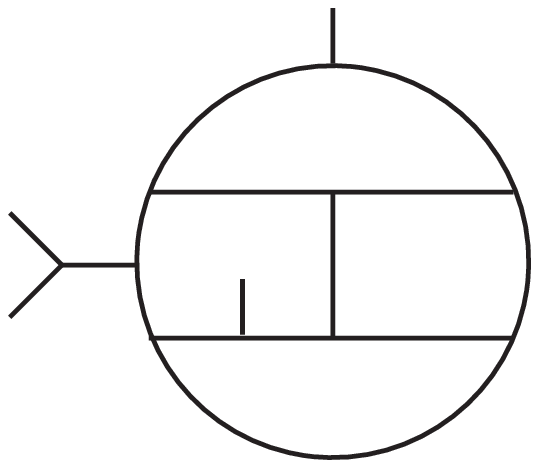}}
\hspace{0.6cm}
\subfigure[]{\hspace{-0.3cm}\includegraphics[clip,scale=0.32]{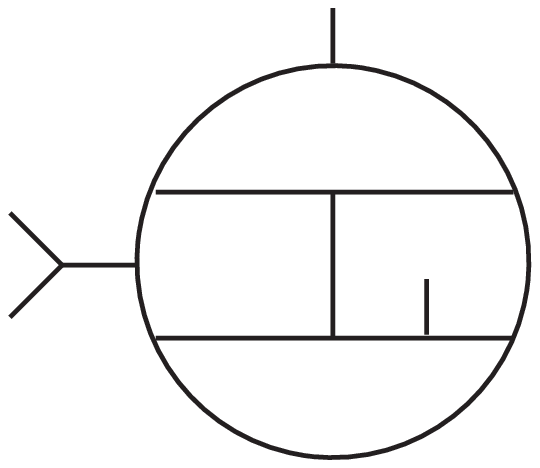}}
\hspace{0.7cm}
\subfigure[]{\hspace{-0.3cm}\includegraphics[clip,scale=0.32]{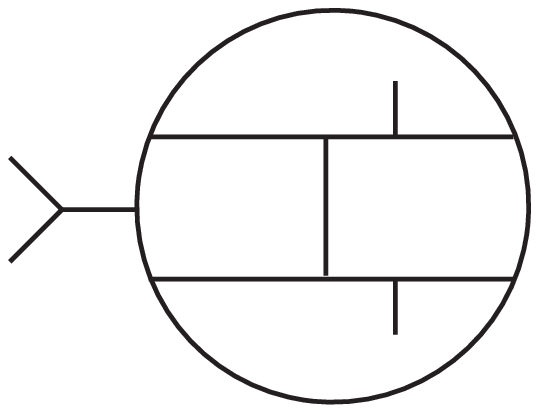}}
\hspace{0.7cm}
\subfigure[]{\hspace{-0.1cm}\includegraphics[clip,scale=0.32]{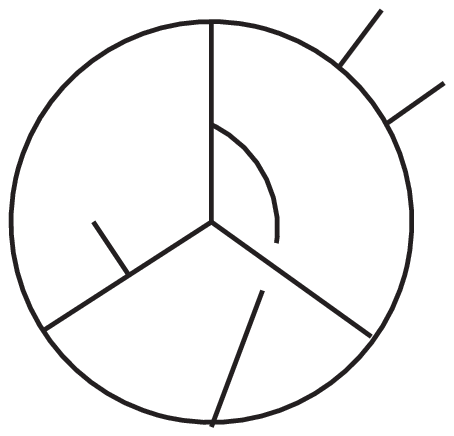}}
\nonumber\\
&
\hspace{0.2cm}
\subfigure[]{\hspace{-0.1cm}\includegraphics[clip,scale=0.32]{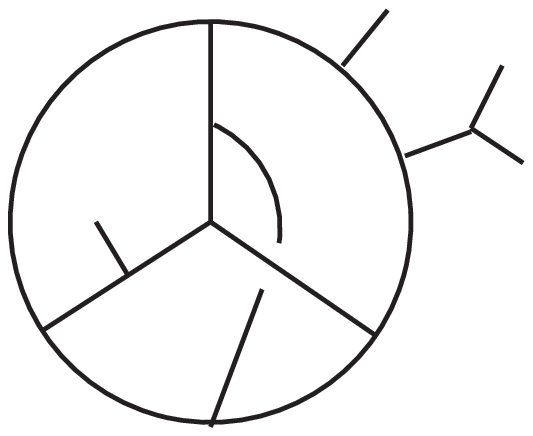}}
\hspace{0.6cm}
\subfigure[]{\hspace{-0.1cm}\includegraphics[clip,scale=0.32]{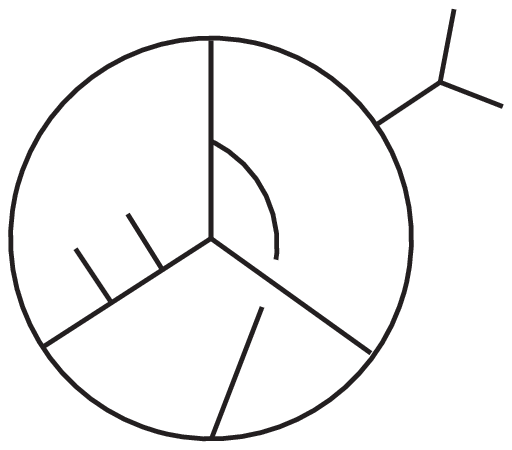}}
\hspace{0.5cm}
\subfigure[]{\hspace{-0.2cm}\includegraphics[clip,scale=0.32]{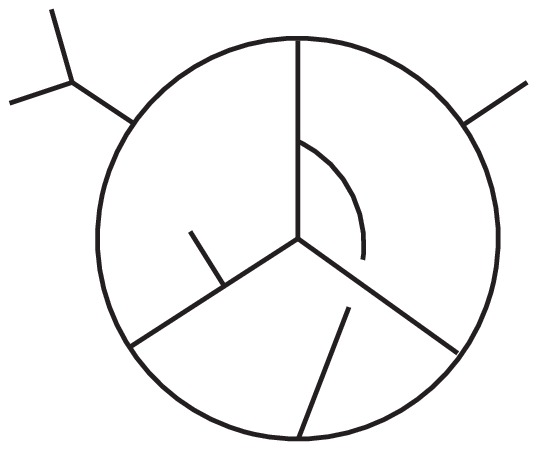}}
\hspace{0.5cm}
\subfigure[]{\hspace{-0.1cm}\includegraphics[clip,scale=0.32]{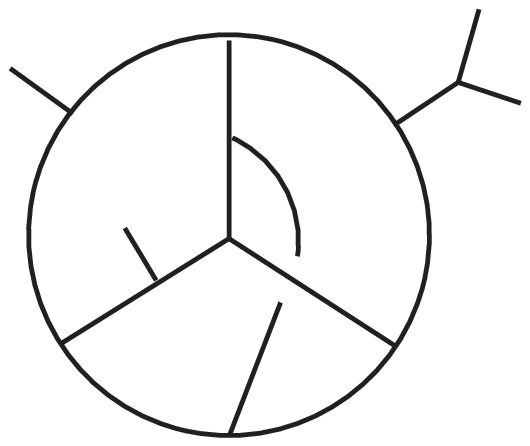}}
\hspace{0.6 cm}
\subfigure[]{\hspace{-0.1cm}\includegraphics[clip,scale=0.32]{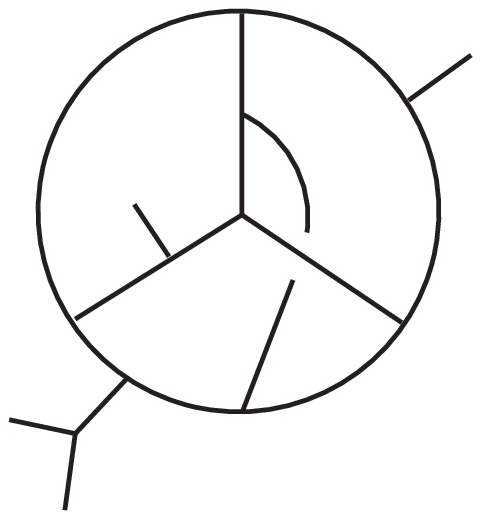}}
\hspace{0.6cm}
\subfigure[]{\hspace{-0.1cm}\includegraphics[clip,scale=0.32]{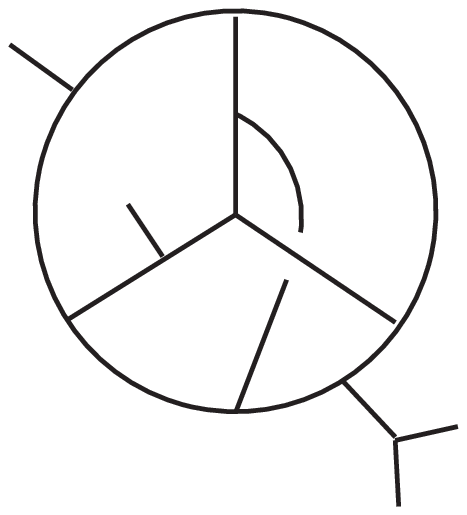}}
\nonumber\\
&
\hspace{1.7cm}
\subfigure[]{\hspace{-0.1cm}\includegraphics[clip,scale=0.32]{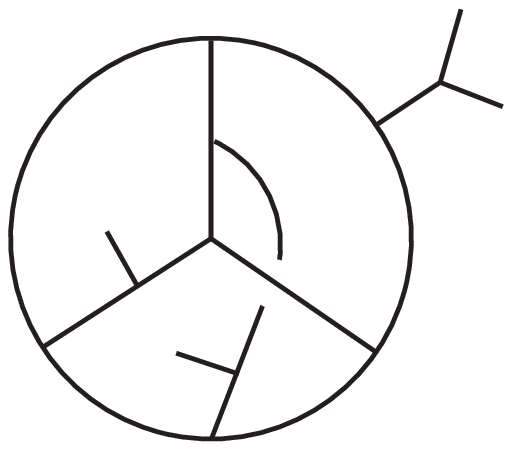}}
\hspace{0.4cm}
\subfigure[]{\hspace{-0.1cm}\includegraphics[clip,scale=0.32]{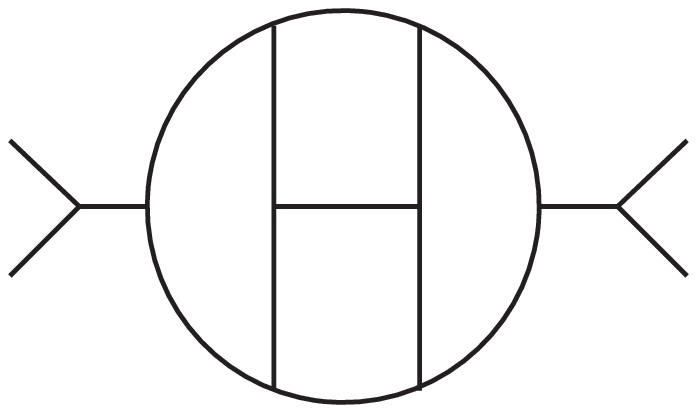}}
\hspace{0.5cm}
\subfigure[]{\hspace{-0.1cm}\includegraphics[clip,scale=0.32]{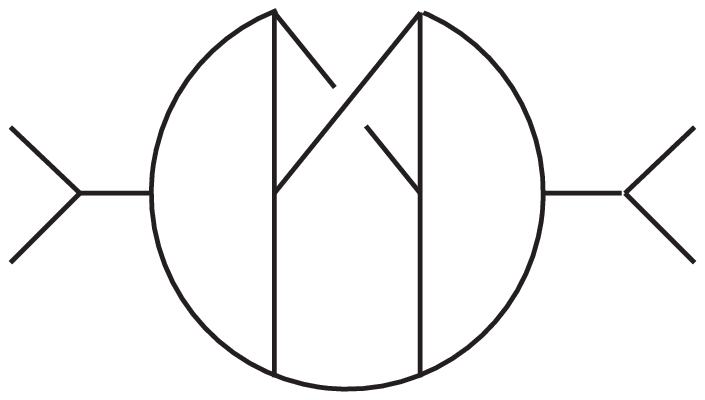}}
\hspace{0.4cm}
\subfigure[]{\hspace{-0.1cm}\includegraphics[clip,scale=0.32]{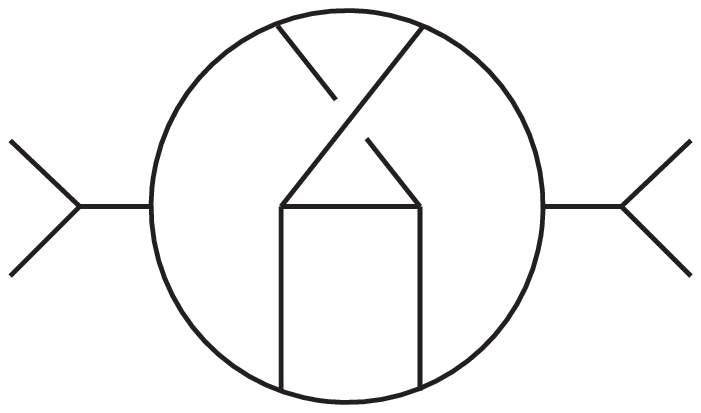}}
\nonumber
\end{align}
\caption{Diagrams 43--82 for the four-loop four-point amplitudes of $\NeqFour$ 
and $\NeqFive$ supergravity. }
\label{FourLoop2Figure}
\end{figure}

\begin{figure}[tbh]
\centering
\renewcommand{\thesubfigure}{(\arabic{subfigure})}
\renewcommand{\subfigcapskip}{-.1cm}
\renewcommand{\subfigbottomskip}{-.1cm}
\setcounter{subfigure}{82}
\subfigure[]{\includegraphics[clip,scale=0.32]{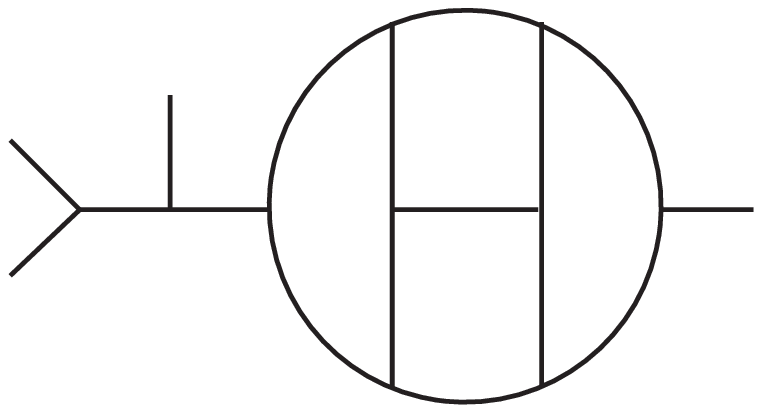}}
\hspace{0.4 cm}
\subfigure[]{\includegraphics[clip,scale=0.32]{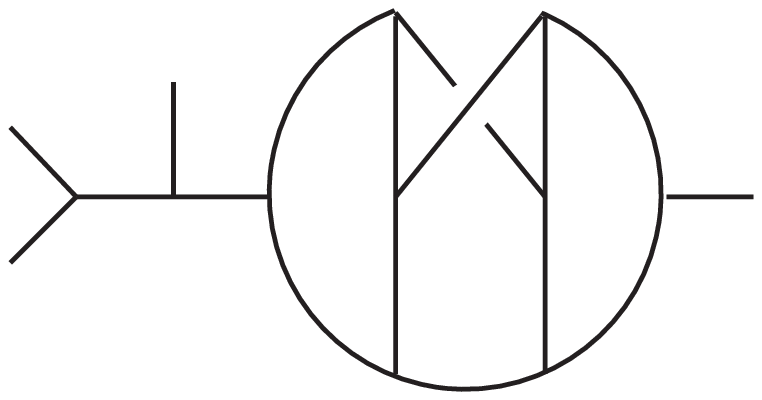}}
\hspace{0.4cm}
\subfigure[]{\includegraphics[clip,scale=0.32]{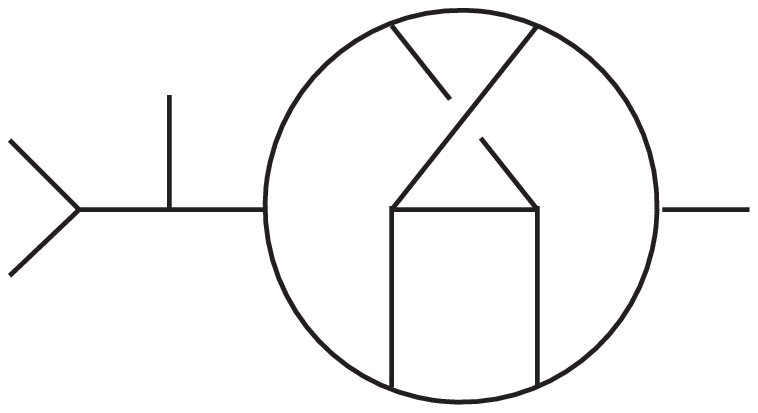}}
\caption{The bubble-on-external-leg diagrams of $\NeqFour$
  super-Yang-Mills theory.  These do not contribute to $\NeqFour$ and
  $\NeqFive$ supergravity.}
\label{FourLoop3Figure}
\end{figure}

\subsection{Review of $\NeqFour$ supergravity}

The calculation of the $\NeqFour$ supergravity divergence starts from
pure Yang-Mills Feynman diagrams, keeping only those diagrams with
color factors that match the 82 $\NeqFour$ super-Yang-Mills diagrams
displayed in \Figs{FourLoop1Figure}{FourLoop2Figure}.
The color factors are then replaced the BCJ forms of the
$\NeqFour$ super-Yang-Mills numerators given in Ref.~\cite{ck4l}.
There are three additional diagrams, displayed in
\Fig{FourLoop3Figure}.  In pure Yang-Mills they contain ultraviolet
divergences (canceled by infrared divergences), but in ${\cal N} \ge
4$ supergravity an extra power of zero in the form of an on-shell
massless external momentum squared in the $\NeqFour$ super-Yang-Mills
numerators sets such potential ultraviolet contributions to zero.

%
\begin{table}[t]
\begin{center}
\scalebox{.9}{
\begin{tabular}[t]{|c|c|}
\hline
graphs & $(\text{divergence}) \times u (4 \pi)^8/(-\langle 12 \rangle^2
[34]^2 st A^\tree (\frac{\kappa}{2})^{10})$ \\
\hline
\hline
1--30 &  \begin{tabular}{@{}c@{}}
$\vspace{.1cm} \vphantom{\Big|^A_B} 
-\!\frac{1}{\eps^4} \Bigl[
 \frac{297863}{3981312}  
s^2+ \frac{7115179}{7962624}  s 
t+ \frac{1230523}{2654208}  t^2
\Bigr] \null +
\frac{1}{\eps^3} \Bigl[
 \frac{183507269}{318504960}  
s^2- \frac{121097629}{106168320}  s 
t- \frac{125340203}{159252480}  t^2
\Bigr]$ \\ 
 $ 
\vspace{0.10cm} 
\null +
\frac{1}{\eps^2} \Bigl[
\zeta_3 \left(- \frac{54780317}{3686400}  
s^2- \frac{364821169}{22118400}  s 
t- \frac{19297919}{7372800}  t^2\right)
\null -
\zeta_2 \left( \frac{297863}{1990656}  
s^2 + \frac{7115179}{3981312}  s 
t+ \frac{1230523}{1327104}  t^2\right)
 \vphantom{\Big|} 
$ \\ 
$ 
\null -
{\rm  S2} \left( \frac{1602535}{73728}s^2
+ \frac{10330175}{442368}  s t
 - \frac{14079343}{442368}  t^2\right)
\null - \frac{80222068879}{28665446400}  
s^2- \frac{949461174731}{57330892800}  s 
t- \frac{17877740021}{19110297600}  t^2
\Bigr]$ \\
$ \vspace{0.10cm}   
\null +
\frac{1}{\eps^{\null}} \Bigl[
\zeta_5 \left( \frac{42165713}{92160}  
s^2+ \frac{12876011}{9216}  s 
t+ \frac{10040753}{46080}  t^2\right)
\null +
\zeta_4 \left( \frac{1162609}{7372800}  
s^2+ \frac{183267071}{14745600}  s 
t+ \frac{110749763}{14745600}  t^2\right)
$\\
$  \vspace{0.10cm}
\null -
\zeta_3 \left( \frac{10506518408983}{71663616000} s^2
 + \frac{30289233413171}{71663616000}  s t 
- \frac{2013863213191}{35831808000}  t^2\right)
\null -
\zeta_2 \left( \frac{970317931}{159252480} s^2 
+ \frac{59367181}{5898240}  s t 
\right.
 \vphantom{\Big|} 
$\\
$ \vspace{0.10cm}
\left.
\null 
- \frac{719420377}{79626240}  t^2\right)
\null -
{\rm T1ep} \left( \frac{1602535}{995328} s^2 
+ \frac{10330175}{5971968}  s t
- \frac{14079343}{5971968}  t^2\right)
\null -
{\rm  S2} \left( \frac{33354691993}{53084160} s^2 \right.
 \vphantom{\Big|} 
$ \\
$ \vspace{0.10cm}   
 \left. \null
+ \frac{19386147397}{10616832}  s t
+ \frac{9723954001}{8847360}  t^2\right) 
 \vphantom{\Big|} 
\null -
{\rm D6}  \left( \frac{4137589}{552960} s^2 
+ \frac{2283701}{184320}  s t
+ \frac{527011}{138240}  t^2\right)
 \vphantom{\Big|} 
$ \\
$ \vspace{0.10cm}   
\null 
- \frac{20252328329611}{143327232000}  
s^2- \frac{534679988685821}{1146617856000}  s 
t- \frac{8363829769903}{1146617856000}  t^2
 \vphantom{\Big|} 
\Bigr] 
$
\end{tabular}
\\ 
\hline
31--60 &  \begin{tabular}{@{}c@{}}
$\vspace{.10cm} 
\vphantom{\Big|^A_B }
\null \frac{1}{\eps^4} \Bigl[
 \frac{1788617}{3981312}  
s^2+ \frac{20728021}{7962624}  s 
t+ \frac{2452169}{2654208}  t^2
\Bigr]
\null +
\frac{1}{\eps^3} \Bigl[
 \frac{527762531}{318504960}  
s^2+ \frac{1120727089}{106168320}  s 
t+ \frac{122147731}{53084160}  t^2
\Bigr]$ \\ 
$\vspace{0.10cm}  
\null +
\frac{1}{\eps^2} \Bigl[
\zeta_3 \left( \frac{6081287}{345600}  
s^2+ \frac{13983243}{819200}  s 
t+ \frac{98182043}{22118400}  t^2\right)
\null + 
  \zeta_2 \left( \frac{1788617}{1990656}  
s^2+ \frac{20728021}{3981312}  s 
t+ \frac{2452169}{1327104}  t^2\right)
 \vphantom{\Big|} 
$ \\
 $ \vspace{0.10cm}
\null +
{\rm  S2} \left( \frac{3516907}{73728}  
s^2+ \frac{31188941}{442368}  s 
t- \frac{15998365}{442368}  t^2\right)
\null +
 \frac{545203990507}{28665446400}  
s^2+ \frac{4109230335503}{57330892800}  s 
t+ \frac{142686680113}{19110297600}  t^2
\Bigr]$ \\  
$ \vspace{0.10cm} 
\null +
\frac{1}{\eps^{\null}} \Bigl[
\zeta_5 \left(- \frac{160438583}{245760}  
s^2- \frac{311758955}{147456}  s 
t- \frac{119748949}{368640}  t^2\right)
\null
 - \zeta_4 \left( \frac{5925797}{921600}  
s^2+ \frac{460780679}{14745600}  s t 
+ \frac{126445477}{14745600}  t^2\right)
 \vphantom{\Big|} 
$ \\ 
$ \vspace{0.10cm} 
\null +
\zeta_3 \left( \frac{11662905491459}{53747712000}  
s^2+ \frac{54035183618969}{71663616000}  s 
t+ \frac{8467395805631}{214990848000}  t^2\right)
\null +
\zeta_2 \left( \frac{3059935571}{159252480} s^2
+ \frac{789428243}{17694720}  s t \right.
 \vphantom{\Big|}   
$ \\
$ \vspace{0.10cm}   
\left.
\null - \frac{197819569}{26542080}  t^2\right)
\null +
{\rm T1ep} \left( \frac{3516907}{995328}  
s^2+ \frac{31188941}{5971968}  s 
t- \frac{15998365}{5971968}  t^2\right)
\null +
{\rm  S2} \left( \frac{2658637313}{53084160} s^2
\right.
 \vphantom{\Big|} 
$ \\ 
$ \vspace{0.10cm} 
\left.
\null
+ \frac{2611873009}{10616832}  s t
+ \frac{23301734753}{26542080}  t^2\right)
\null +
{\rm D6}  \left( \frac{6050189}{552960}  
s^2+ \frac{10479103}{552960}  s 
t+ \frac{233987}{46080}  t^2\right)
\vphantom{\Big|}                                                               
$ \\
$ \vspace{0.10cm}
\null +
\frac{455464156513}{1911029760}  
s^2+ \frac{173334911330293}{229323571200}  s 
t+ \frac{673760034799}{25480396800}  t^2
 \vphantom{\Big|} 
\Bigr] 
$
\end{tabular}
\\ 
\hline
61--82 &  \begin{tabular}{@{}c@{}}
$\vspace{.1cm} \vphantom{\Big|^A_B }
-\! \frac{1}{\eps^4} \Bigl[
\frac{248459}{663552}  
s^2 + \frac{756269}{442368}  s 
t + \frac{610823}{1327104}  t^2
\Bigr]
\null +
\frac{1}{\eps^3} \Bigl[
- \frac{17781745}{7962624}  
s^2- \frac{5553497}{589824}  s 
t- \frac{24110299}{15925248}  t^2
\Bigr]$ \\
 $ \vspace{0.10cm}     
\null +
\frac{1}{\eps^2} \Bigl[
\zeta_3 \left(- \frac{30260233}{11059200} s^2
- \frac{1590799}{2764800}  s t 
- \frac{20144143}{11059200}  t^2\right)
\null -
\zeta_2 \left( \frac{248459}{331776} s^2 
+ \frac{756269}{221184}  s t 
+ \frac{610823}{663552}  t^2\right)
 \vphantom{\Big|} 
$ \\ 
$ \vspace{0.10cm} 
\null -
{\rm  S2} \left( \frac{53177}{2048} s^2 
+ \frac{3476461}{73728}  s t 
-  \frac{319837}{73728}  t^2\right)
- \frac{38748493469}{2388787200}  
s^2- \frac{9752373953}{176947200}  s 
t- \frac{31202235023}{4777574400}  t^2
\Bigr]$ \\ 
 $\vspace{0.10cm} 
\null +
\frac{1}{\eps^{\null}} \Bigl[
\zeta_5 \left( \frac{28798009}{147456}  
s^2+ \frac{35247593}{49152}  s 
t+ \frac{876065}{8192}  t^2\right)
\null +
\zeta_4 \left( \frac{15414589}{2457600}  
s^2+ \frac{11563067}{614400}  s 
t+ \frac{7847857}{7372800}  t^2\right)
$ \\ 
$\vspace{0.10cm} 
 \vphantom{\Big|} 
\null -
\zeta_3 
\left( \frac{15920366514887}{214990848000} s^2
 +  \frac{4001452799633}{11943936000}  s t 
+ \frac{20550575084777}{214990848000}  t^2\right)
\null -
\zeta_2 \left( \frac{52240441}{3981312} s^2
+ \frac{30566335}{884736}  s t \right.
$ \\
$\vspace{0.10cm} \left.
\null
+ \frac{12596167}{7962624}  t^2\right)
\null -
{\rm T1ep} \left( \frac{53177}{27648}  s^2 
+ \frac{3476461}{995328}  s t 
- \frac{319837}{995328}  t^2\right)
\null +
{\rm  S2} \left( \frac{767401367}{1327104} s^2 \right.
$ \\
 $\vspace{0.10cm}
\left. \null 
+ \frac{1397856199}{884736}  s t
+ \frac{587012725}{2654208}  t^2\right)
\null -
{\rm D6}  \left( \frac{47815}{13824} s^2 
+ \frac{22675}{3456}  s t 
+ \frac{17495}{13824}  t^2\right)
$ \\
 $\vspace{0.10cm}
\null
- \frac{434546648527}{4478976000}  
s^2- \frac{9221628964379}{31850496000}  s 
t- \frac{5488842949013}{286654464000}  t^2
 \vphantom{\Big|} 
\Bigr] 
$
\end{tabular}
\\ 
\hline
\hline
sum &  \begin{tabular}{@{}c@{}}
$ \vphantom{\Big|}
\frac{1}{\eps^{\null}} s u \frac{1}{72}(264 \, \zeta_3 - 1)
$
\end{tabular}
\\ 
\hline
\end{tabular}
}
\end{center}
\caption{The divergence in the four-graviton amplitude of pure
  $\NeqFour$ supergravity.  The first three entries correspond to the
  sum over diagrams 1--30, 31--60 and 61--82 listed in
  \figs{FourLoop1Figure}{FourLoop2Figure}, while the final row gives
  the sum over all diagrams. Subdivergences automatically cancel
  amongst themselves and are not included.  Our choice of external
  helicity states and reference momenta are as in
  \tab{HalfMaxUnsubTable}.}
\label{TableN4FourLoop}
\end{table}

After feeding the integrand so constructed through the integration
procedure summarized in \sect{MethodsSection}, we find that pure
$\NeqFour$ supergravity in $D=4$ is divergent~\cite{FourLoopN4}:
\begin{equation}
\mathcal{M}_4^\fourloop\Bigr|_{\mathcal{N}=4,\,\mathrm{div.}} = \frac{1}{(4\pi)^8} \frac{1}{\eps} 
\left(\frac{\kappa}{2}\right)^{10} \frac{1}{144} (1 -264\, \zeta_3)
 \,\mathcal{T} \,,
\label{SupergravityDivergence}
\end{equation}
where $\eps = (4-D)/2$ is the dimensional-regularization parameter, and
\begin{equation}
\mathcal{T} =  s t A_{\NeqFour}^{\tree} 
\left(\mathcal{O}_1-28\mathcal{O}_2-6\mathcal{O}_3\right)\,,
\label{DivergentTensor}
\end{equation}
with
\begin{align}
\mathcal{O}_1& = \sum_{\mathcal{S}_4} \left(D_{\alpha}F_{1\mu\nu}\right)
\left(D^{\alpha}F_2^{\mu\nu}\right)F_{3\rho\sigma}F_4^{\rho\sigma}\,, \notag \\
\mathcal{O}_2& = \sum_{\mathcal{S}_4} \left(D_{\alpha}F_{1\mu\nu}\right)
\left(D^{\alpha}F_2^{\nu\sigma}\right)F_{3\sigma\rho}F_4^{\rho\mu}\,, \\
\mathcal{O}_3& = \sum_{\mathcal{S}_4} \left(D_{\alpha}F_{1\mu\nu}\right)
\left(D_{\beta}F_2^{\mu\nu}\right)F_{3\sigma}^{\hphantom{\sigma}\alpha}
  F_4^{\sigma\beta}\,. \notag
\end{align}
The sum runs over all 24 permutations of the external legs.  
$F_j^{\mu\nu}$ is the
linearized field-strength tensor given in terms of
polarization vectors for leg $j$ as 
\begin{align}
F_j^{\mu\nu}&\equiv i(k_j^{\mu}\varepsilon_j^{\nu}
-k_j^{\nu}\varepsilon_j^{\mu})\,, \notag \\
D^{\alpha}F_j^{\mu\nu}&\equiv -k_j^{\alpha}
(k_j^{\mu}\varepsilon_j^{\nu}-k_j^{\nu}\varepsilon_j^{\mu})\,.
\label{FieldStrength}
\end{align}
This form makes explicit the fact that each state of pure $\NeqFour$ 
supergravity corresponds to a direct product of a
color-stripped state of $\NeqFour$ super-Yang-Mills theory and of pure
nonsupersymmetric Yang-Mills theory.

By taking linear combinations, the divergences can be separated into
distinct helicity classifications:
\begin{align}
\mathcal{O}^{--++}&=\mathcal{O}_1-4\mathcal{O}_2\,, \hskip .8 cm 
\mathcal{O}^{-+++}=\mathcal{O}_1-4\mathcal{O}_3\,,\notag \\
\mathcal{O}^{++++}&=\mathcal{O}_2\,. 
\label{HelicityOperators}
\end{align}
Each of the obtained operators are nonvanishing only for the indicated
helicity configurations and their parity conjugates and
relabelings. The helicity labels refer to those of the polarization
vectors used in \eqn{FieldStrength} on the pure Yang-Mills side and
not the supergravity states, which are obtained by a direct product of
these states with those of $\NeqFour$ super-Yang-Mills theory.  For
explicit helicity states in $D=4$, we have
\begin{align}
\mathcal{O}^{--++} &= 4 s^2 t \frac{\spa1.2^4}{\spa1.2\spa2.3\spa3.4\spa4.1}\,,
                      \notag\\
\mathcal{O}^{-+++} &= -12 s^2 t^2 
                 \frac{\spb2.4^2}{\spb1.2\spa2.3\spa3.4\spb4.1}\,,
                  \label{Operators}   \notag \\
\mathcal{O}^{++++} &= 3 s t (s+t) \frac{\spb1.2\spb3.4}{\spa1.2\spa3.4}\,,
\end{align}
using four-dimensional spinor-helicity notation.

As explained in Ref.~\cite{FourLoopN4}, the appearance of the
divergences in the three independent helicity configurations in
\eqn{HelicityOperators} is unexpected and points to the source of the
divergence being the Marcus $U(1)$ duality-symmetry anomaly~\cite{MarcusAnomaly}.  Without
the anomaly, the ${-}{+}{+}{+}$ and ${+}{+}{+}{+}$ helicity sectors would
vanish.  Ref.~\cite{RaduAnomaly} explains how the anomaly leads to
poor ultraviolet behavior even in the ${-}{-}{+}{+}$ sector.

In \tab{TableN4FourLoop} we have collected together groups of diagrams
in order to display the nontrivial cancellations between diagrams.
The first three entries correspond to the sums over diagrams 1--30,
31--60 and 61--82, while the final one gives the sum over all
diagrams. The final sum displays an enormous cancellation between the
diagrams to yield a remarkably simple result. We do not include
subdivergences which automatically cancel amongst themselves.  In the
table, the $\zeta_i$ are the standard Riemann zeta constants. The
value of ${\rm S2}$ is already defined in \eqn{S2Def}, while the other
constants are~\cite{Czakon}
\begin{align}
{\rm T1ep} & = -\frac{45}{2} - \frac{\pi \sqrt{3} \log^23}{8}
- \frac{35 \pi^3 \sqrt{3}}{216} - \frac{9}{2} \zeta_2 + \zeta_3 
+ 6 \sqrt{3} \, {\rm Cl}_2\left( \frac{\pi}{3} \right) \nonumber \\
& \hskip 2 cm 
- 6 \sqrt{3} \, {\rm Im} \left( {\rm Li}_3 
          \left(\frac{e^{-i \pi/6}}{\sqrt{3}} \right) \right) \,, \nonumber\\
{\rm D6} & = 6 \zeta_3 - 17 \zeta_4 - 4 \zeta_2 \log^2 2 
           + \frac{2}{3} \log^4 2 + 16 {\rm Li}_4 \left(\frac{1}{2} \right) 
           - 4 \left({\rm Cl}_2 \left(\frac{\pi}{3} \right) \right)^2 \,.
\end{align}
As noted earlier, ${\rm Cl}_2(x) = {\rm Im}({\rm Li}_2(e^{ix}))$ is the
Clausen function.
These transcendental constants arise from our use of an infrared
mass regulator and, as expected, cancel from the final
ultraviolet divergence~\cite{ChetyrkinUniformMass,ChetyrkinThreeLoop}.

\begin{table}[t]
\begin{center}
\scalebox{.9}{
\begin{tabular}[t]{|c|c|}
\hline
graphs & $(\text{divergence}) \times u (4 \pi)^8/(-\langle 12 \rangle^2
[34]^2 st A^\tree (\frac{\kappa}{2})^{10})$ \\
\hline
\hline
1--30 &  \begin{tabular}{@{}c@{}}
$\vspace{0.1cm} 
\vphantom{\Big|^A_B }
\frac{1}{\eps^4} \Bigl[
 \frac{607}{1990656}  
s^2- \frac{1323773}{1990656}  s 
t- \frac{14255}{41472}  t^2
\Bigr]  
\null +
\frac{1}{\eps^3} \Bigl[
 \frac{4865671}{19906560} s^2
+ \frac{149977}{3317760}  s t
- \frac{20170049}{19906560}  t^2
\Bigr]$ \\ 
 $ \vspace{0.10cm}   
\null +
\frac{1}{\eps^2} \Bigl[
\null 
\zeta_3 \left(- \frac{3733153}{230400} s^2
- \frac{5900609}{276480}  s t+ 
\frac{3883097}{691200}  t^2\right)
\null +
\zeta_2 \left( \frac{607}{995328}  
s^2- \frac{1323773}{995328}  s 
t- \frac{14255}{20736}  t^2\right)
 \vphantom{\Big|} 
$ \\
 $ \vspace{0.10cm}
\null -
{\rm  S2} \left( \frac{625357}{36864} s^2
+ \frac{5161189}{110592}  s t 
- \frac{1428583}{55296}  t^2\right)
\null 
- \frac{7648139167}{3583180800}  
s^2- \frac{22568882383}{3583180800}  s 
t- \frac{55681241}{59719680}  t^2
\Bigr]$ \\ 
 $ \vspace{0.10cm} 
\null +
\frac{1}{\eps^{\null}} \Bigl[
 \vphantom{\Big|} 
\zeta_5 \left(- \frac{225641}{1024}  
s^2- \frac{12931021}{18432}  s 
t- \frac{2378855}{18432}  t^2\right)
\null -
\zeta_4 \left( \frac{4044329}{460800} s^2
+ \frac{3646153}{921600}  s t
- \frac{2056603}{153600}  t^2\right)
\vphantom{\Big|}                                                               
$ \\
 $ \vspace{0.10cm} 
\null +
\zeta_3 \left( \frac{6076575618157}{17915904000} s^2
+ \frac{3396579085657}{3583180800}  s t
+ \frac{2089036585637}{8957952000}  t^2\right)
\null 
-\zeta_2 \left( \frac{51416459}{9953280} s^2 
+ \frac{801749}{51840}  s t \right.
 \vphantom{\Big|} 
$ \\
 $ \vspace{0.10cm} 
\left.\null
- \frac{65544931}{9953280}  t^2\right)
\null -
{\rm T1ep} \left( \frac{625357}{497664} s^2
 + \frac{5161189}{1492992}  s t 
- \frac{1428583}{746496}  t^2\right)
\null -
{\rm  S2} \left( \frac{8055438013}{16588800} s^2
\right.
 \vphantom{\Big|} 
$ \\
 $ \vspace{0.10cm} 
\left.\null
+ \frac{555755309}{414720}  s t 
+ \frac{555207793}{614400}  t^2\right)
\null -
{\rm D6}  \left( \frac{715513}{138240} s^2
+ \frac{718247}{76800}  s t
+ \frac{285839}{172800}  t^2\right)
 \vphantom{\Big|} 
$ \\
 $ \vspace{0.10cm} 
\null +
 \frac{1916368326173}{71663616000} s^2
+ \frac{7258817218703}{71663616000}  s t
+ \frac{3175133834231}{35831808000}  t^2
\Bigr] 
 \vphantom{\Big|}
$
\end{tabular}
\\ 
\hline
31--60 &  \begin{tabular}{@{}c@{}}
$\vspace{0.1cm} 
\vphantom{\Big|^A_B } 
\frac{1}{\eps^4} \Bigl[
 \frac{509381}{1990656}  
s^2+ \frac{3991391}{1990656}  s 
t+ \frac{242555}{331776}  t^2
\Bigr]
+ \frac{1}{\eps^3} \Bigl[
 \frac{50554927}{19906560}  
s^2+ \frac{13023425}{1327104}  s 
t+ \frac{8356667}{3317760}  t^2
\Bigr]$ \\
  $ \vspace{0.10cm} 
\null +
\frac{1}{\eps^2} \Bigl[
\zeta_2 \left( \frac{509381}{995328}  
s^2+ \frac{3991391}{995328}  s 
t+ \frac{242555}{165888}  t^2\right)
\null +
\zeta_3 \left( \frac{990949}{57600}  
s^2+ \frac{570691}{1382400}  s 
t- \frac{10906963}{691200}  t^2\right)
 \vphantom{\Big|} 
$ \\
 $ \vspace{0.10cm} 
\null +
{\rm  S2} \left( \frac{1380997}{36864}  
s^2+ \frac{9202651}{110592}  s 
t- \frac{821453}{27648}  t^2\right)
\null +
 \frac{65553264229}{3583180800}  
s^2+ \frac{27992599379}{447897600}  s 
t+ \frac{12366245939}{1194393600}  t^2
\Bigr]$ \\ 
 $ \vspace{0.10cm} 
\null +
\frac{1}{\eps^{\null}} \Bigl[
\null
\zeta_5 \left( \frac{10240481}{23040}  
s^2+ \frac{96847583}{92160}  s 
t+ \frac{3535453}{30720}  t^2\right)
\null +
\zeta_4 \left( \frac{816643}{76800}  
s^2- \frac{6008467}{307200}  s 
t- \frac{51227}{2048}  t^2\right)
 \vphantom{\Big|} 
$ \\
$ \vspace{0.10cm} 
\null -
\zeta_3 \left( \frac{8235182625383}{13436928000} s^2
+ \frac{25298224196579}{17915904000}  s t
+ \frac{11561841643253}{53747712000}  t^2\right)
\null +
\zeta_2 \left( \frac{174844657}{9953280} s^2
+ \frac{31428727}{663552} s t \right.
 \vphantom{\Big|} 
$ \\
$ \vspace{0.10cm} 
\left.
- \frac{8072393}{1658880} t^2\right)
\null +  
{\rm T1ep} \left( \frac{1380997}{497664}  
s^2+ \frac{9202651}{1492992}  s 
t- \frac{821453}{373248}  t^2\right)
\null +
{\rm  S2} \left( \frac{2385329963}{16588800}  s^2 \right.
 \vphantom{\Big|} 
$ \\
 $ \vspace{0.10cm} 
\left.\null
+ \frac{1077896293}{3317760}  s t
+ \frac{1501624967}{2073600}  t^2\right)
\null +
{\rm D6}  \left( \frac{233051}{46080}  
s^2+ \frac{4649023}{691200}  s 
t+ \frac{77389}{172800}  t^2\right)
 \vphantom{\Big|} 
$ \\
 $ \vspace{0.10cm} 
\null 
- \frac{273686499733}{2654208000}  
s^2- \frac{10212410685517}{35831808000}  s 
t- \frac{501121685203}{4777574400}  t^2
\Bigr] 
 \vphantom{\Big|} 
$
\end{tabular}
\\ 
\hline
61--82 &  \begin{tabular}{@{}c@{}}
$\vspace{0.1cm} 
\vphantom{\Big|^A_B }
- \!\frac{1}{\eps^4} \Bigl[
 \frac{42499}{165888}  
s^2 + \frac{148201}{110592}  s 
t+ \frac{128515}{331776}  t^2
\Bigr]
\null +
\frac{1}{\eps^3} \Bigl[
- \frac{27710299}{9953280}  
s^2- \frac{21805693}{2211840}  s 
t- \frac{29969953}{19906560}  t^2
\Bigr]$ \\
  $ \vspace{0.10cm} 
\null +
\frac{1}{\eps^2} \Bigl[
\zeta_3 \left(- \frac{25627}{25600}  
s^2+ \frac{1607353}{76800}  s 
t+ \frac{3511933}{345600}  t^2\right)
\null +
\zeta_2 \left(- \frac{42499}{82944}  
s^2- \frac{148201}{55296}  s 
t- \frac{128515}{165888}  t^2\right)
 \vphantom{\Big|} 
$ \\
 $ \vspace{0.10cm} 
\null +
{\rm  S2} \left(- \frac{10495}{512}  
s^2- \frac{673577}{18432}  s 
t+ \frac{71441}{18432}  t^2\right)
\null 
- \frac{9650854177}{597196800}  
s^2- \frac{7458218987}{132710400}  s 
t- \frac{11252621119}{1194393600}  t^2
\Bigr]
 \vphantom{\Big|} $ \\
  $ \vspace{0.10cm} 
\null +
\frac{1}{\eps^{\null}} \Bigl[
\zeta_5 \left(- \frac{10327117}{46080}  
s^2- \frac{1788471}{5120}  s 
t+ \frac{321979}{23040}  t^2\right)
\null +
\zeta_4 \left(- \frac{855529}{460800}  
s^2+ \frac{10835777}{460800}  s 
t+ \frac{892711}{76800}  t^2\right)
 \vphantom{\Big|} 
$ \\
 $ \vspace{0.10cm} 
\null +
\zeta_3 \left( \frac{14908078591061}{53747712000}  
s^2+ \frac{1396836736049}{2985984000}  s 
t- \frac{972377870569}{53747712000}  t^2\right)
\null +
\zeta_2 \left(- \frac{61714099}{4976640}  
s^2- \frac{35277233}{1105920}  s t
\right.
 \vphantom{\Big|} 
$ \\
 $ \vspace{0.10cm} 
\left.\null
- \frac{17110573}{9953280}  t^2\right)
\null +
{\rm T1ep} \left(- \frac{10495}{6912}  
s^2- \frac{673577}{248832}  s 
t+ \frac{71441}{248832}  t^2\right)
\null +
{\rm  S2} \left( \frac{113402161}{331776} s^2 \right.
 \vphantom{\Big|} 
$ \\
 $ \vspace{0.10cm} 
\left.\null
+ \frac{1122715393}{1105920}  s t
+ \frac{119104427}{663552}  t^2\right)
\null + 
{\rm D6}  \left( \frac{409}{3456}  
s^2+ \frac{2269}{864}  s 
t+ \frac{4169}{3456}  t^2\right)
 \vphantom{\Big|} 
$ \\
 $ \vspace{0.10cm} 
\null +
 \frac{2736085919309}{35831808000}  
s^2+ \frac{1462778758259}{7962624000}  s 
t+ \frac{1166557609583}{71663616000}  t^2
\Bigr] 
$
\end{tabular}
\\ 
\hline
\hline
sum &  \begin{tabular}{@{}c@{}}
$\vspace{0.10cm} \vphantom{\Big|}
- \!\frac{1}{\eps^{\null}} s u  \frac{1}{72}  \left(264 \, \zeta_3 -1 \right)
$
\end{tabular}
\\
\hline
\end{tabular}
}
\end{center}
\caption{The additional contributions in $\NeqFive$ supergravity.
  These include internal states that arise from a direct product of
  the $\NeqFour$ sYM states and a Majorana fermion. The sum of these
  contributions together with the ones in \tab{TableN4FourLoop}
  vanishes, showing that the $\NeqFive$ supergravity amplitude is
  ultraviolet finite.  Subdivergences automatically cancel amongst
  themselves and are not included.  Our choice of external helicity
  states and reference momenta are as in \tab{HalfMaxUnsubTable}.  }
\label{TableNfFourLoop}
\end{table}

\subsection{$\NeqFive$ supergravity}

Next we turn to $\NeqFive$ supergravity in $D=4$.  As discussed in
\sect{N5ConstructionSubsection}, we obtain $\NeqFive$ supergravity
from $\NeqFour$ supergravity by adding in the contributions from the
BCJ construction based on the direct product of $\NeqFour$
super-Yang-Mills with additional contributions from adding a single
Majorana fermion to the pure gluon theory.  The calculation is
somewhat more complicated than the pure $\NeqFour$ supergravity
calculation because of the long fermion traces that appear at four loops.

Because $\NeqFive$ supergravity has no duality-symmetry anomaly, we
expect it to be ultraviolet finite at four loops.  Indeed, additional
$\NeqOne$ supersymmetry identities are sufficient to show amplitudes
and any associated potential ultraviolet divergences vanish in the
${-}{+}{+}{+}$ and ${+}{+}{+}{+}$ helicity sectors in \eqn{Operators}.
Only the ${-}{-}{+}{+}$ sector gives nonvanishing amplitudes and
therefore needs checking, although we have calculated the other two
sectors as well.  We have only computed the case where the external
$\NeqFive$ supergravity states are those obtained from a direct
product of $\NeqFour$ super-Yang-Mills and pure Yang-Mills states
(i.e. the subset of states that are also in the pure $\NeqFour$
supergravity spectrum).  However, as mentioned in
\sect{ThreeLoopSection}, $\NeqOne$ and $\NeqFour$ supersymmetry
identities in the direct product~\cite{SWI,TwoLoopGluino} allow us to
express any of the four-point amplitudes in terms of one of them, so
ruling out divergences in this sector rules out {\it all} four-point
divergences.

We find that the four-loop four-point amplitudes of
$\NeqFive$ supergravity are finite.  This may be unsurprising given
that the additional supersymmetry compared to the $\NeqFour$
supergravity case should improve the ultraviolet properties.  However,
the fact remains that, at present, there is no standard-symmetry
explanation for the finiteness.  In addition, as we explain in
\sect{PowerCountingSection}, no covariant diagrammatic formalism can
display the cancellations manifestly, so the vanishing of the
divergence is another example of enhanced cancellations.

In \tab{TableNfFourLoop} we give the extra contributions to 
the potential divergence coming from
the additional states that are present in $\NeqFive$ supergravity
compared to $\NeqFour$ supergravity.  As can be seen from the final
entry in the table, the contribution is equal and opposite to the
contribution that comes solely from the $\NeqFour$ supergravity states
given in \tab{TableN4FourLoop}.  Therefore the total divergence
vanishes:
\begin{equation}
\mathcal{M}_4^\fourloop\Bigr|_{\mathcal{N}=5,\,\mathrm{div.}} = 0\,.
\end{equation}
 The nontrivial way the
cancellations occur in the sum of the entries in
\tabs{TableN4FourLoop}{TableNfFourLoop} suggests that there should be
a better way to see them.  While it may be simple to state the
obvious, as already explained in \sect{PowerCountingSection}, finding
a formalism that makes these cancellations manifest is nontrivial,
given that no covariant diagrammatic representation exists that does
so.

Besides the information given in \tabs{TableN4FourLoop}
{TableNfFourLoop}, in accompanying Mathematica
attachments~\cite{AttachedFile} we give the divergences for each
diagram for pure $\NeqFour$ supergravity as well as for the additional
contributions needed for $\NeqFive$ supergravity.  As in the tables,
we do not include subdivergences in these files since they
automatically cancel amongst themselves in our calculation.

We note that while our calculation proves that there are no four-loop
four-point divergences in $\NeqFive$ supergravity, it does not rule
out five-point $R^5$-type divergences.  It would of course be
interesting to study these as well in the future.

\section{Conclusions}
\label{Conclusions}

In this paper, we described the phenomenon of {\it enhanced
  ultraviolet cancellations} in supergravity theories.  By definition,
when all local covariant diagrammatic representations of an amplitude
contain terms that have a worse power count than the amplitude as a
whole, we have enhanced cancellations.  To illustrate this phenomenon,
we first discussed $\NeqFour$ supergravity in four dimensions at three
loops.  By power counting maximal cuts, we identified terms in the
four-point amplitude that are divergent at three loops, in agreement
with supersymmetry and duality-symmetry
arguments~\cite{VanishingVolume}, when, in fact, the amplitude is
three-loop finite in $D=4$~\cite{ThreeLoopN4}.  The theory does diverge at four
loops~\cite{FourLoopN4}, but it appears to be due to a rigid $U(1)$
duality-symmetry anomaly~\cite{MarcusAnomaly, RaduAnomaly}.  Such
anomalies are not present in $\N \ge 5$ supergravity theories,
suggesting that these theories cannot have similar divergences.  This
motivated us to study the four-loop four-point amplitudes of
$\NeqFive$ supergravity.  Again in this case, power counting maximal
cuts identifies divergent terms in $D=4$, consistent with
standard-symmetry considerations~\cite{VanishingVolume}.  However,
explicit calculations performed in this paper shows this amplitude is
ultraviolet finite, again illustrating enhanced ultraviolet
cancellations.  If similar enhanced ultraviolet cancellations hold in
$\NeqEight$ supergravity, then this theory will be finite at seven
loops as well, contradicting predictions based on power
counting~\cite{SevenLoopGravity, BjornssonGreen, VanishingVolume}.

The underlying reason for enhanced ultraviolet cancellations is not
fully understood.  There are some indications that the duality between
color and kinematics~\cite{BCJ,BCJLoop} is responsible.  An explicit
study shows that this duality is responsible for improved ultraviolet
behavior in the relatively simple two-loop case of half-maximal
supergravity in $D=5$~\cite{HalfMax5D}.  An important challenge is to
push this understanding to higher loop orders.  Another important
question is whether there might be an explanation for enhanced
cancellations based on supersymmetry or duality symmetry.  Such an
explanation would have to be novel, given that enhanced ultraviolet
cancellations are nonstandard.  The potential three-loop
$R^4$ counterterm of $\NeqFour$ supergravity and four-loop counterterm
${\cal D}^2 R^4$ counterterm of $\NeqFive$ supergravity cannot be
written as full superspace integrals~\cite{VanishingVolume}.  An
interesting open question is whether this plays any role in the vanishing
of the associated divergences. 

While we do not yet know if perturbatively ultraviolet-finite unitary
field theories of gravity exist, based on the results of this paper it
is clearly premature to conclude otherwise.  More generally,
nontrivial multiloop enhanced cancellations in gravity theories
are a new and surprising phenomenon, contrary to expectations based on
standard-symmetry considerations which suggest viable counterterms.
The existence of these cancellations gives us confidence that further
nontrivial surprises await us as we probe supergravity theories to
ever higher loop orders using modern tools.

\subsection*{Acknowledgments}
We thank Guillaume Bossard, John Joseph Carrasco, Paolo Di Vecchia,
Sergio Ferrara, Enrico Herrmann, Ian Jack, Henrik Johansson, Tim
Jones, Sean Litsey, Josh Nohle, Radu Roiban, James Stankowicz, Kelly
Stelle and Jaroslav Trnka for helpful discussions.  We especially
thank Alexander and Volodya Smirnov for their assistance with the
integrals.  This material is based upon work supported by the
Department of Energy under Award Number DE-{S}C0009937.  We also
thank the Danish Council for Independent Research. We gratefully
acknowledge Mani Bhaumik for his generous support.  We also thank
Academic Technology Services at UCLA for computer support.


\end{document}